\documentclass[preprint,showpacs,preprintnumbers,amsmath,amssymb]{revtex4}
\usepackage{graphics}  

\begin{document}  

\title{The effect of unitary noise \\ on Grover's quantum search algorithm}

\author{Daniel Shapira}
\author{Shay Mozes}
\author{Ofer Biham}
\affiliation{Racah Institute of Physics, The Hebrew University, Jerusalem,91904,Israel}

\newpage

\begin{abstract}
The effect of unitary noise on the performance of Grover's quantum search 
algorithm is studied. 
This type of noise may result from tiny fluctuations and drift in 
the parameters of the (quantum) components performing the computation.
The resulting operations are still unitary, but not precisely those
assumed in the design of the algorithm. 
Here we focus on the effect of such noise in the Hadamard gate $W$, 
which is an essential component in each iteration of the
quantum search process. 
To this end $W$ is replaced by a noisy Hadamard gate $U$. 
The parameters of $U$ at each iteration are taken from an arbitrary  
probability distribution (e.g. Gaussian distribution) and are characterized 
by their statistical moments around the parameters of $W$. 
For simplicity we assume that the noise is unbiased and isotropic, namely
all noise variables in the parametrization we use have zero average and the
same standard deviation $\epsilon$. 
The noise terms at different calls to $U$ are assumed to be uncorrelated. 
For a search space of size
$N=2^n$ (where $n$ is the number of qubits used to span this space)
it is found that as long as $\epsilon < O(n^{-\frac{1}{2}} N^{-\frac{1}{4}})$,
the algorithm maintains significant efficiency, while above this noise
level its operation is hampered completely.
It is also found that below this noise threshold, when the search fails, it
is likely to provide a state that differs from the marked state by only a few 
bits. This feature can be used to search for the marked state by a classical
post-processing, even if the quantum search has failed, thus improving the success
rate of the search process.
\end{abstract}

\pacs{PACS: 03.67.Lx, 89.70.+c}

\maketitle

\section{Introduction}

The discovery of quantum algorithms that can solve computational 
problems faster than any known classical algorithm stimulated
much interest in quantum information science.
The known algorithms include Shor's factoring algorithm \cite{Shor94,Ekert96a},
Grover's search algorithm  \cite{Grover96,Grover97a} 
as well as algorithms for the simulation of physical systems.
One of the most serious obstacles that should be dealt with in the
way to construct a quantum computer on which such algorithms can
be implemented is the problem of decoherence  \cite{Zurek01}.
This is the effect of the interaction between the quantum system 
that stores and manipulates the quantum information and the 
environment, that spoils the coherence of the quantum states.
The effect of decoherence on Grover's search algorithm was
recently studied using perturbation theory
\cite{Azuma2002}.
Recent progress in quantum error correcting codes  
\cite{Shor95,Shor96a,Steane96,Steane96a,Knill97} as well as 
in decoherence free sub-spaces  \cite{Zanardi97,Lidar98,Bacon99,Kempe01}, 
may provide an effective way to
keep the quantum states coherent and enable the implementation of
useful quantum algorithms. However, a drawback of these
approaches is that they involve some redundancy in the
encoding of the logical quantum state, thus requiring to maintain
a larger number of quantum bits in a coherent state.

The performance of a quantum computer may also be affected by unitary noise 
\cite{Bernstein97,Preskill,Nielsen00}.
Such noise may result from tiny fluctuations
and drift in the properties of the implemented quantum gates.
These fluctuations or drifts may add stochastic perturbation elements 
to the Hamiltonian that describes the quantum gates that  generate the  unitary operations   
in the algorithm.  The perturbated Hamiltonian is Hermitian as well, therefore  
the resulting operations are still unitary, although not precisely the
ones assumed in the design of the algorithm.
Since the implementation of a useful quantum algorithm requires a
large number of one and two qubit gates, 
it is possible that even a tiny noise in each operation would accumulate
to a considerable effect that may hamper the operation of the quantum 
computer. 

In this paper we analyze the effect of unitary noise on Grover's 
quantum search algorithm. 
To this end we replace the Hadamard gate $W$ by a noisy (but still unitary) Hadamard 
gate $U$. The parameters of $U$ at each iteration are taken from an arbitrary probability  
distribution (e.g. Gaussian distribution) and are characterized by their  statistical moments 
around those of $W$. 

In order to simplify the calculations we assume an unbiased noise (i.e. noise with 
mean $0$ ) or alternately we refer to a  known bias that is shifted away in every iteration 
of the Hadamard gate. We assume that the noise is isotropic, namely the standard deviation 
$\epsilon$ is the same for all the noise variables in the parameters to be defined later. 
This assumption is made for simplicity, and removing it does not change the main 
qualitative features observed in this paper. 

We show that the noise reduces the success probability $P_0$ of the algorithm (the probability 
to measure the marked state, the one we are looking for, at the end of the algorithm). 
Consider a search problem with $N$ elements where $N = 2^n$ ($n$ denotes the number of 
qubits in the register). We find that as long as
the standard deviation $\epsilon$ of the noise satisfies

\begin{equation}
\epsilon \alt \epsilon_0 = 1.4 \frac{1}{\sqrt{n \sqrt{N}}} 
\label{large_epsilon}
\end{equation}
the algorithm maintains significant efficiency, while above this noise level, its operation is 
hampered completely.

We have analyzed the flow of probability out of the marked state which is caused by the noise.  
In the effective region of the algorithm (i.e. $\epsilon \alt \epsilon_0 $ ), we find that a 
considerably large amount of probability diffuses from the marked state to its near 
neighbors, whose indices differ only in a few bits from the marked index. This result is 
derived analytically and verified numerically. 

The "diffusive" flow from the marked state to its near neighbors can be used to enhance the 
efficiency of the algorithm. To this end we execute the Grover quantum search, and measure 
the state of the register. If it is not the marked state we {\it classically } test all its 
neighboring states, namely those that differ from it by only a few bits. 
The use of hybrid (quantum and classical) search strategies in case of 
unbiased and isotropic unitary noise in the Hadamard operations accomplishes a great 
reduction in the average searching time in comparison to the ordinary 
quantum search procedure that is 
re-execution of Grover's algorithm over and over again until the marked state is found. 
Moreover, hybrid strategies enable an efficient quantum search under noise levels 
for which the quantum search alone fails. 

The paper is organized as follows: In section II we introduce the Grover quantum search 
algorithm. In section III we analyze the effect of unitary noisy Hadamard gates  on 
the Grover quantum search. In section IV we present and discuss numerical results that 
verify the analytical predictions and extend the scope of discussion to noise's limits for which 
the analytical approximations are not valid anymore. In section V we examine the use 
of hybrid search strategies which improves the performance of the Grover quantum 
search with noisy Hadamard gates. Some additional details of calculations are presented 
in the appendix. 
 
\section{Grover's Search Algorithm}

Let $ D $ be a search space containing $ N $ elements. We
assume, for convenience, that $ N=2^{n} $, where $ n $ is an
integer. In this way, we may represent the elements of $ D $ using
an $ n $-qubit register containing their indices, $ i=0,\dots ,N-1 $.
We assume that a single marked element is the solution to the search
problem. The distinction between the marked and unmarked elements
can be expressed by a suitable function, $ f:D\rightarrow \{0,1\} $,
such that $ f=1 $ for the marked element, and $ f=0 $ for all
other elements. 

Suppose we wish to search the space $ D $ to find the marked element,
namely the element for which $ f=1 $. To solve this problem on
a classical computer one needs to evaluate $ f $ for each element,
one by one, until the marked state is found. Thus, $ O(N) $ evaluations
of $ f $ are required on a classical computer. However, if we allow
the function $ f $ to be evaluated \emph{coherently}, there exists
a sequence of unitary operations which can locate the marked element
using only $ O(\sqrt{N}) $ queries of $ f $ \cite{Grover96,Grover97a}.
This sequence of unitary operations is called Grover's quantum search
algorithm. 

To describe the operation of the quantum search algorithm we first
introduce a register, $ |\bar{x}\rangle =|x_{n-1},x_{n-2},\ldots ,x_{1}, x_{0}\rangle  $
of $ n $ qubits, and an \emph{ancilla} qubit, $ |q\rangle  $,
to be used in the computation. The bar script denotes a computational
string of $ n $ bits, i.e. $ \bar{x}=x_{n-1},x_{n-2},\ldots ,x_{1}, x_{0} $
where the $k$'th bit $ x_{k} $ is either one or zero, so that
$ \left| \bar{x}\right\rangle  $ is a computational basis state
vector. We also introduce a \emph{quantum oracle}, a unitary operator
$ \hat{O}_{f} $ which functions as a black box with the ability
to \emph{recognize} the solution to the search problem. We shall use
the hat script for operators, namely an operator $ O $ is denoted
as $ \hat{O} $. (For more details on how an oracle may be constructed,
see Chapter 6 of Ref. \cite{Nielsen00}.) The oracle performs the
following unitary operation on computational basis states of the register
$ |\bar{x}\rangle  $ and of the ancilla $ |q\rangle  $:

\begin{equation}
\label{eq:bborac}
\hat{O}_{f}|\bar{x}\rangle |q\rangle =|\bar{x}\rangle |q\oplus f(\bar{x})\rangle ,
\end{equation}

\noindent where $ \oplus  $ denotes addition modulo 2. This definition
may be uniquely extended, via linearity, to all states of the register
and ancilla. 

The oracle recognizes the marked state in the sense that if $ \bar{x} $
is the marked element of the search space, $ f(\bar{x})=1 $, the
oracle flips the ancilla qubit from $ |0\rangle  $ to $ |1\rangle  $
and vice versa, while for unmarked states the ancilla is unchanged.
In Grover's algorithm the ancilla qubit is initially set to the state

\begin{equation}
|q\rangle =\frac{1}{\sqrt{2}}\left( |0\rangle -|1\rangle \right) .
\end{equation}

\noindent It is easy to verify that, with this choice, the action
of the oracle is:

\begin{equation}
\hat{O}_{f}|\bar{x}\rangle \left( \frac{|0\rangle -|1\rangle }{\sqrt{2}}\right) =(-1)^{f(\bar{x})}|\bar{x}\rangle \left( \frac{|0\rangle -|1\rangle }{\sqrt{2}}\right) .
\end{equation}

\noindent Thus, the only effect of the oracle is to apply a phase
rotation of $ \pi  $ radians if $ \bar{x} $ is the marked state,
and no phase change if $ \bar{x} $ is unmarked. Since the state
of the ancilla does not change, it is conventional to omit it, and
write the action of the oracle as

\begin{equation}
\hat{O}_{f}|\bar{x}\rangle =(-1)^{f(\bar{x})}|\bar{x}\rangle .
\end{equation}

\noindent Grover's search algorithm may be summarized as follows: 

\begin{enumerate}
\item Initialize the qubit register to $ |\bar{0}\rangle =|0,0,\ldots ,0\rangle  $
and the ancilla to $ |0\rangle  $. Then apply the gate $ \hat{w}\hat{x} $
on the ancilla qubit where $ \hat{x}=\left( \begin{array}{cc}
0 & 1\\
1 & 0
\end{array}\right)  $ is the \textsc{not} gate and $ \hat{w}=\frac{1}{\sqrt{2}}\left( \begin{array}{cc}
1 & 1\\
1 & -1
\end{array}\right)  $ is the Hadamard gate (The matrices are written with respect to the
computational basis $ |0\rangle ,|1\rangle  $). The resulting state
is: \begin{equation}
\label{init:state}
|0,0,\ldots ,0\rangle \left( \frac{|0\rangle -|1\rangle }{\sqrt{2}}\right) _{q}
\end{equation}

\item Grover Iterations: Repeat the following operation $ T $ times: 

\begin{enumerate}
\item \label{en:rot1} Apply the Hadamard gate on each qubit in the register. 
\item \label{en:rot2} Apply the oracle, which has the effect of rotating
the marked state by a phase of $ \pi  $ radians. Since the ancilla
is always in the state $ (|0\rangle -|1\rangle )/\sqrt{2} $ the
effect of this operation may be described by a unitary operator acting
only on the register, 
$ {\hat{I}}_{m}=\sum _{\bar{x}}(-1)^{f(\bar{x})}|\bar{x}\rangle \langle \bar{x}| 
= \hat{I} - 2|\bar{m}\rangle\langle\bar{m}|$. ($\hat{I}$ denotes the $2^n \times 2^n$ 
identity operator and $|\bar{m}\rangle$ denotes the marked state, which is the one we are 
searching for.)
\item \label{en:rot3} Apply the Hadamard gate on each qubit in the register.
\item \label{en:rot4} Rotate the $ |0,0,\ldots ,0\rangle  $ state of
the register by a phase of $ \pi  $. This rotation is similar to
2(b), except for the fact that here it is performed on a known state.
It takes the form 
$ \hat{I}_{0}=-|\bar{0}\rangle \langle \bar{0}|+\sum _{\bar{x}\neq 0}|\bar{x}\rangle \langle \bar{x}| 
=  \hat{I} - 2|\bar{0}\rangle\langle\bar{0}|$. 
\end{enumerate}
\item Apply the Hadamard gate on each qubit in the register and then measure
the register in the computational basis. 
\end{enumerate}
We still need to specify the number of iterations, $ T $. As subsequent
Grover iterations are applied, the amplitude of the marked state gradually
increases, while the amplitudes of the unmarked states decrease. There
exists an optimal number, $ T $, of iterations at which the amplitude
of the marked state reaches a maximum value, and thus the probability
that the measurement yields the marked state is maximal. Let us denote
this probability by $ P_0 $. It has been shown \cite{Grover97a,Boyer96,Zalka99}
that the optimal time $ T $ is:
\begin{equation}
\label{eq:optit}
T\leq \left\lceil \frac{\pi }{4}\sqrt{N}\right\rceil ,
\end{equation}
where $\lceil x \rceil$ denotes the integer value of $x$. 
Moreover, it was shown \cite{Boyer96} that Grover's algorithm is
optimal in the sense that it is as efficient as theoretically possible
\cite{Bennett97}. For the initial state given in
step 1 above, the probability to obtain a marked state at the optimal
time $ T$ is $ P_0=1-O(1/\sqrt{N}) $ \cite{Boyer96,Zalka99}.
For other initial states $ P_0 $ may be bounded away from unity \cite{BihamE99}.
Further generalizations to the original Grover's algorithm have also been suggested 
in \cite{Grover98a,BihamE01a}.

Following Grover's discoveries,  
a variety of applications were developed, in which the
algorithm is used in the solution of other problems 
\cite{Durr96,Grover97b,Grover97c,Terhal97,Brassard98,Grover98a,
Cerf00,Gingrich00,Grover00a,Carlini99};
Experimental implementations were also constructed  using a
nuclear magnetic resonance 
(NMR) quantum computer \cite{Chuang98b,Jones98} 
as well as on an optical device \cite{Kwiat00}.

\section{Analysis of the effects of unitary noise}
Assuming a single marked state, the original Grover quantum search algorithm 
using an $n$-qubit register with $N = 2^n$ computational basis states, could be represented 
by the operation: 
\begin{equation}
| g(T) \rangle  \equiv  {\hat {G}}_0 (T) | \bar{0} \rangle 
= \hat{W} {\hat{Q}}^T  | \bar{0} \rangle.
\label{ket_gT}
\end{equation}     
$| g(T) \rangle $ is the  quantum state after $T$ iterations  which is accomplished  by 
executing the Grover operator ${\hat {G}}_0 (T)$ on the initial state 
$| \bar{0} \rangle  =  | 0 \ldots 00 \rangle$. 
When the  time $T$ is optimal, the operator 
$ {\hat {G}}_0 (T) $ in Eq. (\ref{ket_gT})  
amplifies the absolute value of the amplitude of the marked state to $1$.  
The Grover operator  ${\hat {G}}_0 (T)$ consists of 
$T$ executions of Grover's iteration $\hat{Q}$ followed by a single Hadamard 
transform $\hat{W}$.  The Grover iteration is defined as the composite operation: 
\begin{equation}
\hat{Q} = -{\hat{I}}_0 \hat{W} {\hat{I}}_m \hat{W} 
\label{hat_Q}
\end{equation}
where:
\begin{equation}
\hat{W} = \bigotimes_{k=0}^{n-1} {\hat{w}}_k 
              \equiv \bigotimes_{k=0}^{n-1} \frac{1}{\sqrt{2}} 
             {\left( \begin{array}{cc} 1 & 1 \\ 1 & -1 \end{array} \right)}_k
\label{hat_W}
\end{equation}
is the $n$-qubit Hadamard transform ($\otimes$ denotes a tensor product 
and $\bigotimes_{k=0}^{n-1} {\hat{w}}_k  = \hat{w}_{n-1} \otimes \ldots \otimes \hat{w}_0$), 
$\hat{I}_0=\hat{I}-2 | \bar{0} \rangle \langle \bar{0} | $ 
is a selective rotation by $\pi$ of the state  $| \bar{0} \rangle$ and  
 $\hat{I}_m=\hat{I}-2 | \bar{m} \rangle \langle \bar{m} |$ 
is a selective rotation by $\pi$ of  
the marked state $| \bar{m} \rangle$  ($\hat{I}$  denotes the identity operator).  
The matrix notation used is in the computational basis representation, 
i.e., ${| 0 \rangle}_k = {\left( \begin{array}{c} 1 \\ 0 \end{array} \right)}_k$ and 
${| 1 \rangle}_k = {\left( \begin{array}{c} 0 \\ 1 \end{array} \right)}_k$.  

Grover's iterations consist of the Hadamard transforms ${\hat{w}}_k$, $k = 0, \ldots n-1$ which 
are one qubit gates, and the selective inversion operators that involve two qubit 
gates. Here we consider the effect of unitary noise in the Hadamard gates 
${\hat{w}}_k$, $k = 0, \ldots n-1$ on the performance of Grover's algorithm. 
To this end we define the noisy Grover iteration at time $t$ as: 
\begin{equation}
\hat{Q}_t = -{\hat{I}}_0 {\hat{V}}_t  {\hat{I}}_m {\hat{U}}_t.
\label{hat_Qt}
\end{equation} 
Here ${\hat{U}}_t $ and  ${\hat{V}}_t $  are the direct products of one qubit unitary operators, 
which are built of noisy executions of one qubit Hadamard gates.  We assume that 
there are no temporal correlations in the Hadamard executions and that  operations on 
different qubits are uncorrelated as well. Therefore, the operators can be written as:
\begin{eqnarray}
{\hat{U}}_t  & = & \bigotimes_{k=0}^{n-1} {\hat{w}}_k e^{i {\hat{a}}_k (t) } 
                    \equiv \hat{W} e^{i {\hat{A}}_t} \nonumber \\ 
{\hat{V}}_t & = & \bigotimes_{k=0}^{n-1} e^{i {\hat{c}}_k (t) } {\hat{w}}_k  
                    \equiv e^{i {\hat{C}}_t} \hat{W}
\label{hat_UVt}
\end{eqnarray}
where ${\hat{a}}_k (t)$ and ${\hat{c}}_k (t)$ are one qubit stochastic operators, acting on the 
$k$'th qubit in the register at time $t$. They are Hermitian and generate the unitary noise in the 
one qubit Hadamard gates.  Any one qubit Hermitian operator can be expanded 
in the basis of the Pauli operators ${\hat{\sigma}}_{1k}$, ${\hat{\sigma}}_{2k}$ and  
${\hat{\sigma}}_{3k}$ and the identity operator ${\hat{\sigma}}_{0k}$, acting on the $k$'th qubit. 
Therefore: 
\begin{eqnarray}
{\hat{a}}_k (t) & = & \sum_{\mu = 0}^{3} \alpha_{\mu k} (t) {\hat{\sigma}}_{\mu k} 
\nonumber \\
{\hat{c}}_k (t) & = &  \sum_{\mu = 0}^{3} \gamma_{\mu k} (t) {\hat{\sigma}}_{\mu k} 
\label{hat_act}
\end{eqnarray}
where:
\begin{equation}
{\hat{\sigma}}_{0k}  =  {\left( \begin{array}{cc} 1 & 0 \\ 0 & 1 \end{array} \right)}_k
\hspace{.5in}
{\hat{\sigma}}_{1k}  =  {\left( \begin{array}{cc} 0 & 1 \\ 1 & 0 \end{array} \right)}_k 
\hspace{.5in}
{\hat{\sigma}}_{2k}  =  {\left( \begin{array}{cc} 0 & -i \\ i & 0 \end{array} \right)}_k
\hspace{.5in}
{\hat{\sigma}}_{3k}  =  {\left( \begin{array}{cc} 1 & 0 \\ 0 & -1 \end{array} \right)}_k 
\label{hat:sigma}
\end{equation}
are the identity and Pauli matrices in the computational basis representation and $\alpha_{\mu k} (t) $ 
and $\gamma_{\mu k} (t)$  ($\mu = 0,1,2,3$ and $k = 1, \ldots ,n$) are real stochastic 
variables.  Since all the $\alpha$'s and $\gamma$'s are produced by the same physical 
hardware with no correlations, their statistical properties can be expressed by 
their moments:
\begin{eqnarray} 
\langle \alpha_{\mu k} (t) \rangle  & \equiv &  \langle \alpha_{\mu} \rangle 
\nonumber \\
\langle \gamma_{\mu k} (t) \rangle &  \equiv &  \langle \gamma_{\mu} \rangle 
\label{<alpha_gamma>}
\end{eqnarray}
and 
\begin{eqnarray}
\langle \alpha_{\mu k} (t)  \alpha_{\nu k'} (t') \rangle  & \equiv & 
	\delta_{tt'}\delta_{kk'} \langle \delta \alpha_{\mu} \delta \alpha_{\nu} \rangle  +
	\langle \alpha_{\mu} \rangle  \langle \alpha_{\nu} \rangle 
\nonumber \\
\langle \gamma_{\mu k} (t)  \gamma_{\nu k'} (t') \rangle  & \equiv &
	\delta_{tt'}\delta_{kk'} \langle \delta \gamma_{\mu} \delta \gamma_{\nu} \rangle +
	\langle \gamma_{\mu}\rangle   \langle \gamma_{\nu} \rangle
\label{<alpha2_gamma2>}
\end{eqnarray}
where:
\begin{eqnarray}
\delta \alpha_{\mu} & = & \alpha_{\mu} - \langle \alpha_{\mu} \rangle
\nonumber \\
\delta \gamma_{\mu} & = & \gamma_{\mu} - \langle \gamma_{\mu} \rangle
\label{dalpha_dgamma}
\end{eqnarray}
and $\delta_{tt'}$ as well as $\delta_{kk'}$ are Kronecker's delta functions.  
The noise is thus characterized by the eight real  stochastic variables: $\alpha_{\mu}$ and 
$\gamma_{\mu}$ ($\mu = 0,1,2,3$).

Using a series expansion of the exponentials in Eq. (\ref{hat_UVt}) and the identity 
of operators: 
\[
(\hat{A} \otimes \hat{B})(\hat{C} \otimes \hat{D}) = \hat{A} \hat{C} \otimes \hat{B} \hat{D}
\]
we obtain:
\begin{eqnarray}
\hat{A}_t & = & \sum_{k=0}^{n-1}{\hat{a}}_k (t) \nonumber \\
\hat{C}_t & = & \sum_{k=0}^{n-1}{\hat{c}}_k (t).
\label{hat_ACt}
\end{eqnarray}
The operators ${\hat{A}}_t$ and  ${\hat{C}}_t$ are built as sums of $n$ one qubit 
operators. 
Therefore, the only non-zero elements in their $2^n \times 2^n$ matrix representation in 
the computational basis are those having indices that differ in no more than one bit.
Denoting two computational basis states: 
$| \bar{x} \rangle = | x_{n-1},x_{n-2}, \ldots x_0 \rangle$ and 
$| \bar{y} \rangle = | y_{n-1},y_{n-2}, \ldots y_0 \rangle$ 
($x_k = 0,1$ and $y_k = 0,1$ where $k = 0, \ldots ,n-1$), 
the following matrix elements can be calculated using Eqs. 
(\ref{hat_act})-(\ref{hat:sigma}) and (\ref{hat_ACt}): 
\begin{equation}
\langle \bar{x} |{\hat{A}}_t| \bar{y} \rangle = \left\{ \begin{array}{cc}
	\sum_{k=0}^{n-1} \{\alpha_{0k} (t) + (-1)^{y_k}\alpha_{3k} (t) \}  &
	\mbox{if  $||\bar{x} - \bar{y}|| = 0$ ,}\\
	\alpha_{1k} (t) + i(-1)^{y_k}\alpha_{2k} (t)  &  
	\mbox{if  $||\bar{x} - \bar{y}|| = 1$  $(x_k = \neg y_k)$,} \\
	0 & \mbox{if $||\bar{x} - \bar{y}|| > 1$}
\end{array}
\right. 
\label{<Axy>}
\end{equation}
and 
\begin{equation}
\langle \bar{x} | {\hat{C}}_t| \bar{y} \rangle = \left\{ \begin{array}{cc}
	\sum_{k=0}^{n-1} \{\gamma_{0k} (t) + (-1)^{y_k}\gamma_{3k} (t) \} &
	\mbox{if  $||\bar{x} - \bar{y}|| = 0$ ,}\\ 
	\gamma_{1k} (t) + i(-1)^{y_k}\gamma_{2k} (t)  &  
	\mbox{if  $||\bar{x} - \bar{y}|| = 1$  $(x_k = \neg y_k)$,} \\
	0 & \mbox{if $||\bar{x} - \bar{y}|| > 1$.}
\end{array}
\right.
\label{<Cxy>}
\end{equation}
The norm  $||\bar{x} - \bar{y}||$ is the Hamming distance that counts the number of 
bits that are different in $\bar{x}$ and $\bar{y}$. 
Similarly, one can write:
\begin{eqnarray}
\hat{W} \hat{A}_t \hat{W} & = & \sum_{k=0}^{n-1}{\hat{w}}_k \hat{a}_k (t) {\hat{w}}_k
\nonumber \\
\hat{W} \hat{C}_t \hat{W} & = & \sum_{k=0}^{n-1}{\hat{w}}_k \hat{c}_k (t) {\hat{w}}_k
\label{hat_WACtW}
\end{eqnarray}
so that:
\begin{equation}
\langle \bar{x} | \hat{W} {\hat{A}}_t \hat{W} | \bar{y} \rangle = \left\{ \begin{array}{cc}
	\sum_{k=0}^{n-1} \{\alpha_{0k} (t) + (-1)^{y_k}\alpha_{1k} (t) \} &
	\mbox{if  $||\bar{x} - \bar{y}|| = 0$ ,}\\
	\alpha_{3k} (t) - i(-1)^{y_k}\alpha_{2k} (t)  &  
	\mbox{if  $||\bar{x} - \bar{y}|| = 1$ $(x_k = \neg y_k)$,} \\
	0 & \mbox{if $||\bar{x} - \bar{y}|| > 1$}
\end{array}
\right.
\label{<WAxyW>}
\end{equation}
and 
\begin{equation}
\langle \bar{x} | \hat{W} {\hat{C}}_t \hat{W} | \bar{y} \rangle = \left\{ \begin{array}{cc}
	\sum_{k=0}^{n-1} \{ \gamma_{0k} (t) + (-1)^{y_k}\gamma_{1k} (t) \} &
	\mbox{if  $||\bar{x} - \bar{y}|| = 0$ ,}\\
	 \gamma_{3k} (t) - i(-1)^{y_k}\gamma_{2k} (t)  &  
	\mbox{if  $||\bar{x} - \bar{y}|| = 1$ $(x_k = \neg y_k)$,} \\
	0 & \mbox{if $||\bar{x} - \bar{y}|| > 1$}
\end{array}
\right.
\label{<WCxyW>}
\end{equation}
are elements of  $2^n \times 2^n$ sparse matrices as well. 

The noisy Hadamard transforms ${\hat{U}}_t$ and  $ {\hat{V}}_t$  
defined in Eq. (\ref{hat_UVt}) can now be written explicitly into the noisy Grover's iteration 
(\ref{hat_Qt})  for any time $t$: 
\begin{eqnarray}
{\hat{Q}}_t  & = &
 -{\hat{I}}_0 e^{i {\hat{C}}_t} \hat{W} {\hat{I}}_m \hat{W} e^{i {\hat{A}}_t}  \nonumber \\
& = &
({\hat{I}}_0e^{i{\hat{C}}_t}{\hat{I}}_0)(-{\hat{I}}_0\hat{W}{\hat{I}}_m\hat{W} )e^{i{\hat{A}}_t} \nonumber \\
& = &{\hat{I}_0}e^{i{\hat{C}}_t}{\hat{I}_0}\hat{Q}e^{i{\hat{A}}_t} 
= e^{i\hat{I}_0\hat{C}_t\hat{I}_0}\hat{Q}e^{i{\hat{A}}_t},
\label{Qt} 
\end{eqnarray}
and the entire search process after $T$ iterations is then given by:
\begin{eqnarray}
\hat{G}(T) & = &{\hat{U}}_{T+1} {\hat{Q}}_T \ldots {\hat{Q}}_1 \nonumber \\ 
& = & 
(\hat{W}e^{i{\hat{A}}_{T+1}})(e^{i\hat{I}_0\hat{C}_T\hat{I}_0}\hat{Q}e^{i{\hat{A}}_T})
(e^{i\hat{I}_0\hat{C}_{T-1}\hat{I}_0} \ldots 
e^{i{\hat{A}}_2})(e^{i\hat{I}_0\hat{C}_1\hat{I}_0}\hat{Q}e^{i\hat{A}_1}).
\label{hat_GT}
\end{eqnarray}
A Taylor series expansion now yields 
\begin{eqnarray}
e^{i{\hat{A}}_{t+1}}e^{i\hat{I}_0\hat{C}_t\hat{I}_0} 
& \approx & (\hat{I}+i{\hat{A}}_{t+1}-\frac{1}{2}{\hat{A}_{t+1}}^2 + \ldots )
(\hat{I}+i\hat{I}_0\hat{C}_{t}\hat{I}_0-\frac{1}{2}\hat{I}_0\hat{C}_{t}^2\hat{I}_0 + \ldots)
\nonumber \\
& \approx & 
\hat{I}+i{{\hat{E}}_t}^{(1)}-\frac{1}{2}{{\hat{E}}_t}^{(2)} + \ldots
\label{Et_Taylor}
\end{eqnarray}
where
\begin{eqnarray}
{{\hat{E}}_t}^{(1)} & = & \hat{A}_{t+1} + {\hat{I}}_0 {\hat{C}}_t {\hat{I}}_0 
\nonumber \\
{{\hat{E}}_t}^{(2)} & = & 
{\hat{A}_{t+1}}^2 + 2\hat{A}_{t+1}{\hat{I}}_0 {\hat{C}}_t {\hat{I}}_0  +
{\hat{I}}_0 {{\hat{C}}_t}^2{\hat{I}}_0
\label{hat_Et12}
\end{eqnarray}
are the first and the second order deviations, caused by the noise, so that:
\begin{eqnarray}
\hat{G}(T) & \approx & 
\hat{W}(\hat{I}+i{{\hat{E}}_T}^{(1)}-\frac{1}{2}{{\hat{E}}_T}^{(2)} + \ldots)
\hat{Q}(\hat{I}+i{{\hat{E}}_{T-1}}^{(1)}-\frac{1}{2}{{\hat{E}}_{T-1}}^{(2)} + \ldots)
\ldots 
\hat{Q}(\hat{I}+i{{\hat{E}}_{0}}^{(1)}-\frac{1}{2}{{\hat{E}}_{0}}^{(2)} + \ldots)
\nonumber \\
& =  & {\hat{G}_0}(T) + {\hat{G}_1}(T) + {\hat{G}_2}(T) + \ldots .
\label{hat_GT_approx}
\end{eqnarray}
where we define $\hat{C}_0 \equiv 0$.
The operator 
\begin{equation}
{\hat{G}_0}(T) =  \hat{W} {\hat{Q}}^T 
\label{hat_G0T}
\end{equation}
is the original Grover quantum search operator (without noise), while $\hat{G}_l (T)$ 
are the components of perturbation of  order $l$ ($l = 1,2 \ldots$).  
Specifically, the leading terms of the perturbation satisfy:
\begin{equation}
{\hat{G}_1}(T) = i\hat{W} \{
\sum_{t=0}^{T}({\hat{Q}}^{T-t}{{\hat{E}}_t}^{(1)}{\hat{Q}}^t)\}  
\label{hat_G1T}
\end{equation}
and 
\begin{equation}
{\hat{G}_2}(T) = -\frac{1}{2}\hat{W} \{
\sum_{t=0}^{T} ({\hat{Q}}^{T-t} {{\hat{E}}_t}^{(2)} {\hat{Q}}^t) + 
2\sum_{t=1}^{T} \sum_{t'=0}^{t-1} 
({\hat{Q}}^{T-t} {{\hat{E}}_t}^{(1)} {\hat{Q}}^{t-t'} 
{{\hat{E}}_{t'}}^{(1)} {\hat{Q}}^{t'}) \}.
\label{hat_G2T}
\end{equation}
Those additional perturbation components consist of summations over time indices $t$ of operator 
multiplications. These multiplications include the noise generators 
${\hat{A}}_t$ and  ${\hat{C}}_t$ where the number of their 
appearances determines the order of the perturbation. They also 
include non-stochastic operators: powers of the original Grover iteration ${\hat{Q}}$ and selective 
inversions around the initial state of zeros ${\hat{I}}_0$.  In order to understand the behavior of 
the operator $\hat{G}(T)$, let us first focus on the non-stochastic 
operators which appear in those multiplications.  

For any arbitrary state  $| \psi \rangle$ of the $n$-qubit register 
\begin{eqnarray}
\hat{Q}| \psi \rangle & = & -{\hat{I}}_0 \hat{W} {\hat{I}}_m \hat{W}| \psi \rangle 
 \nonumber \\
& = & 
-(\hat{I}-2| \bar{0} \rangle \langle \bar{0} |)\hat{W}(\hat{I}-
2| \bar{m} \rangle \langle \bar{m} |)\hat{W}| \psi \rangle \nonumber \\ 
& = & 
-| \psi \rangle  + 
2 \{ \langle \bar{0} |  \psi \rangle 
-2\langle \bar{0} | \hat{W}| \bar{m} \rangle \langle \bar{m} | 
\hat{W}| \psi \rangle \}
| \bar{0} \rangle  + 
2 \langle \bar{m} | \hat{W}| \psi \rangle  \hat{W}| \bar{m} \rangle . 
\label{hat_Qpsi}
\end{eqnarray}
In particular:
\begin{equation}
\hat{Q}| \bar{0} \rangle  = \{1-4{|\langle \bar{m} |  \hat{W}| \bar{0} \rangle |}^2 \} 
| \bar{0} \rangle + 2\langle \bar{m} | \hat{W}| \bar{0} \rangle 
 \hat{W}| \bar{m} \rangle 
\label{hat_Q0}
\end{equation} 
and 
\begin{equation}
\hat{Q}\hat{W}| \bar{m} \rangle = 
-2\langle \bar{0} | \hat{W}| \bar{m} \rangle | \bar{0} \rangle  + 
\hat{W}| \bar{m} \rangle.
\label{hat_QWm}
\end{equation}
The operator  $\hat{Q}$ acts as a linear transformation within a 3-dimensional vector space that is 
spanned by the three vectors: $| \psi \rangle $, $| \bar{0} \rangle $ and $\hat{W}| \bar{m} \rangle $. 
In case that the state vector  $| \psi \rangle $ is linearly independent on the vectors  
$| \bar{0} \rangle $ and 
$\hat{W}| \bar{m} \rangle$, the following representation of vectors basis can be performed: 
\begin{equation}
| \psi \rangle \leftrightarrow \left( \begin{array}{c} 1 \\ 0 \\ 0  \end{array} \right)
\hspace{.5in}
| \bar{0} \rangle \leftrightarrow \left( \begin{array}{c} 0 \\ 1 \\ 0 \end{array} \right)
\hspace{.5in}
\hat{W}| \bar{m} \rangle \leftrightarrow \left( \begin{array}{c} 0 \\ 0 \\ 1 \end{array} \right).
\label{3d_vectors}
\end{equation}

Note that the these vectors are  {\it not} represented in the computational basis. 
However, they are linearly independent and hence can be treated as a standard 
basis in a simple 3-dimensional vector space in which an inner product is {\it not} 
defined. Namely,  the above representation does {\it not} imply orthogonality of 
the vectors $| \psi \rangle $, $| \bar{0} \rangle $ and $\hat{W}| \bar{m} \rangle $, 
and the following matrix calculations are exact and involve no approximations at 
all.  
This representation does not change eigenvalues and eigenvectors from
those obtained for an orthonormal basis.

Now, denoting 
$\langle \bar{0} |\hat{W}| \bar{m} \rangle = 
\langle \bar{m} | \hat{W}| \bar{0} \rangle = \frac{1}{\sqrt{N}}$, 
the operator $\hat{Q}$, which is a regular linear transformation acting on a simple 3-dimensional 
vector space, is represented by the 3-dimensional matrix: 
\begin{equation}
\hat{Q} = \left( \begin{array}{ccc} 
-1 & 0 & 0 \\
q_1 & 1 - \frac{4}{N} & -\frac{2}{\sqrt{N}} \\
q_2 & \frac{2}{\sqrt{N}} & 1
\end{array} \right)
\label{Q_matrix}
\end{equation} 
where:
\begin{eqnarray}
q_1 & = & 2 \{\langle \bar{0} |\psi\rangle -
 \frac{2}{\sqrt{N}}\langle \bar{m} | \hat{W}| \psi \rangle  \}
\nonumber \\
q_2 & = & 2 \langle \bar{m} | \hat{W}| \psi \rangle .
\label{q12}
\end{eqnarray}
Powers of the operator $\hat{Q}$ do not exceed from the 3-dimensional sub-space spanned by 
$| \psi \rangle $, $| \bar{0} \rangle $ and $ \hat{W}| \bar{m} \rangle$. 
Diagonalizing the matrix $\hat{Q}$ we obtain 
\begin{equation}
{\hat{Q}}_D = \left( \begin{array}{ccc} 
{\lambda}_0 & 0 & 0 \\
0 & {\lambda}_+ & 0 \\
0 & 0 & {\lambda}_- 
\end{array} \right)
\label{QD_matrix}
\end{equation}
where the eigenvalues are 
$\lambda_0 = -1$
and 
$\lambda_\pm = e^{\pm i \omega}$,
and $\omega$ is given by:
\begin{equation}
\cos \omega = 1 - \frac{2}{N}.  
\label{cosw}
\end{equation}
The diagonalizing matrix whose columns consist of the three eigenvectors takes the form:
\begin{equation}
\hat{X} = \left( \begin{array}{ccc}
1 & 0 & 0 \\
v_1 & 1 & 1 \\
v_2 & e^{+i\varphi} & e^{-i\varphi} 
\end{array} \right)
\label{hat_X}
\end{equation}
where
\begin{equation}
e^{\pm i \varphi} = \frac{-\frac{2}{\sqrt{N}}}{1 - e^{\pm i \omega}} 
\label{eiphi}
\end{equation}
and 
\begin{eqnarray}
v_1 & = & 
\frac{\frac{1}{\sqrt{N}} \langle \bar{m}|\hat{W}{| \psi} \rangle 
-\langle \bar{0} |\psi \rangle}
{1-\frac{1}{N}}
\nonumber \\
v_2 & = & 
\frac{\frac{1}{\sqrt{N}}\langle \bar{0} | \psi \rangle 
- \langle \bar{m} | \hat{W}{| \psi \rangle }}
{1-\frac{1}{N}}.
\label{v1v2}
\end{eqnarray}
We now obtain an explicit expression for powers of  $\hat{Q}$ 
\begin{eqnarray}
{\hat{Q}}^t & = & \hat{X} {\hat{Q}_D}^t {\hat{X}}^{-1} \nonumber \\
& = & \left(  \begin{array}{ccc}
{(-1)}^t & 0 & 0 \\
v_1 [{(-1)}^t-k(t)] + v_2 l(t) & k(t) & -l(t) \\
v_2 [{(-1)}^t-m(t)] - v_1 l(t) & l(t) & m(t) 
\end{array} \right)
\label{Q^t}
\end{eqnarray}
where:
\begin{eqnarray}
k(t) & = & -\frac{\sin(\omega t - \varphi)}{\sin(\varphi)} \nonumber \\
l(t) & = & -\frac{\sin(\omega t)}{\sin(\varphi)} \nonumber \\
m(t) & = & \frac{\sin(\omega t + \varphi)}{\sin(\varphi)}.
\label{klm}
\end{eqnarray}

The expressions above can be simplified in the limit of large $N$.  
In this limit the frequency  $\omega$ given by Eq. (\ref{cosw}) can be approximated by 
\begin{equation}
\omega = \frac{2}{\sqrt{N}} + O \left( \frac{1}{{N}^{\frac{3}{2}}} \right). 
\label{omega}
\end{equation}
Under this approximation  Eq. (\ref{eiphi}) becomes 
\[
e^{\pm i \varphi} = \mp i + O\left(\frac{1}{\sqrt{N}}\right), 
\]
namely
\[
\varphi = -\frac{\pi}{2} + O\left(\frac{1}{\sqrt{N}}\right).   
\]
As a result:
\begin{equation}
k(t) \approx m(t) = \cos (\omega t ) + O\left(\frac{1}{\sqrt{N}}\right)
\label{coswt}
\end{equation}
and
\begin{equation}
l(t) = \sin(\omega t ) + O\left(\frac{1}{\sqrt{N}}\right).
\label{sinwt}
\end{equation}
Moreover, as we shall see later, the superposition $| \psi \rangle $ appears in our first order calculation 
in only two forms.  
Either $| \psi \rangle =| \bar{y} \rangle $ where $| \bar{y} \rangle $ 
is a computational basis state 
perpendicular to $| \bar{0} \rangle $ or $| \psi \rangle =\hat{W}| \bar{y} \rangle $ where 
$| \bar{y} \rangle $ is a computational basis state perpendicular to the marked state 
$| \bar{m} \rangle $. 
In the first case we obtain 
\begin{eqnarray}
v_1 = \frac{\frac{1}{\sqrt{N}}\langle \bar{m} | \hat{W}| \bar{y} \rangle }
{1-\frac{1}{N}} 
& = & O\left(\frac{1}{N}\right)   \nonumber \\
v_2 = \frac{-\langle \bar{m} | \hat{W}| \bar{y} \rangle }{1-\frac{1}{N}} 
& = & O\left(\frac{1}{\sqrt{N}}\right). 
\label{v1v2y}
\end{eqnarray}
In the second case we find that 
\begin{eqnarray}
v_1 = \frac{-\langle \bar{0} | \hat{W}| \bar{y} \rangle }{1-\frac{1}{N}} 
& = & O\left(\frac{1}{\sqrt{N}}\right) \nonumber \\ 
v_2 = \frac{\frac{1}{\sqrt{N}}\langle \bar{0} | \hat{W}| \bar{y} \rangle }
{1-\frac{1}{N}} 
& = & O\left(\frac{1}{N}\right).   
\label{v1v2Wy}
\end{eqnarray}
Thus in both cases the matrix of Eq. (\ref{Q^t}) takes the form:
\begin{equation}
{\hat{Q}}^t =  \left( \begin{array}{ccc} 
{(-1)}^t & 0 & 0 \\
O(\frac{1}{\sqrt{N}})   & \cos{(\omega t)} & -\sin{(\omega t)} \\
O(\frac{1}{\sqrt{N}})  & \sin{(\omega t)} &  \cos{(\omega t)}
\end{array} \right) 
\label{Q^tapprox}
\end{equation}
This means that  ${\hat{Q}}^t$ acts simultaneously, up to $O(\frac{1}{\sqrt{N}})$, as a 
2-dimensional rotator in the sub-space 
spanned by the state vectors  $| \bar{0} \rangle $ and  $\hat{W}| \bar{m} \rangle$,  
as well  as selective phase invertor (according to the parity of the power $t$) of any state vector 
$| \psi \rangle$ that is independent on   $| \bar{0} \rangle $ and  $\hat{W}| \bar{m} \rangle$.  

The selective inversion around $| \bar{0} \rangle$ (denoted as ${\hat{I}}_0$) 
also preserves the 3-dimensional simple vector space   spanned by  $| \psi  \rangle$, 
For any superposition $| \psi \rangle $
\begin{equation}
{\hat{I}}_0 | \psi \rangle = (\hat{I} - 2| \bar{0} \rangle \langle \bar{0} | )
| \psi \rangle = | \psi \rangle  - 2\langle \bar{0} | \psi \rangle | \bar{0} \rangle,
\label{I0|psi>}
\end{equation}
where $\langle\bar{0}|\psi \rangle =0$ for the case of $| \psi \rangle =| \bar{y} \rangle 
 \neq | \bar{0} \rangle$ 
and  $\langle\bar{0}|\psi \rangle  = \frac{1}{\sqrt{N}}$ for the case of 
 $| \psi \rangle =\hat{W}| \bar{y} \rangle$ 
($| \bar{y} \rangle  \neq | \bar{m} \rangle$), 
where $| \bar{y} \rangle$ denotes an arbitrary  computational basis state. Particularly we can write:
 \begin{equation}
{\hat{I}}_0 | \bar{0}  \rangle = - | \bar{0}  \rangle 
\label{I0|0>}  
\end{equation}
and
 \begin{equation}
{\hat{I}}_0 \hat{W} | \bar{m} \rangle =\hat{W} | \bar{m} \rangle  
- \frac{2}{\sqrt{N}} | \bar{0}  \rangle 
\label{I0|Wm>}  
\end{equation}
Hence in both cases the selective inversion can be represented by: 
\begin{equation}
{\hat{I}}_0 = \left( \begin{array}{ccc}
1 & 0 & 0 \\
O(\frac{1}{\sqrt{N}}) & -1 &  O(\frac{1}{\sqrt{N}})  \\
0 & 0 & 1 
\end{array} \right).
\label{I0approx}
\end{equation}

The effect of the stochastic operators ${\hat{A}}_t$ and ${\hat{C}}_t$, given by 
Eqs.  (\ref{<Axy>}) - (\ref{<WCxyW>}) on a computational basis state $| \bar{y} \rangle$ 
is given by:
\begin{eqnarray}
{\hat{A}}_t| \bar{y} \rangle & = & 
\langle\bar{y}|{\hat{A}}_t| \bar{y} \rangle | \bar{y} \rangle + 
\sum_{k=0}^{n-1} 
\langle {\bar{y}}_{1k} | {\hat{A}}_t| \bar{y} \rangle | {\bar{y}}_{1k} \rangle  
\nonumber \\
{\hat{A}}_t \hat{W} | \bar{y} \rangle & = &
\langle\bar{y}| \hat{W} {\hat{A}}_t \hat{W} | \bar{y} \rangle \hat{W} | \bar{y} \rangle + 
\sum_{k=0}^{n-1} 
\langle{\bar{y}}_{1k}| \hat{W} {\hat{A}}_t \hat{W} | \bar{y} \rangle 
\hat{W} | {\bar{y}}_{1k} \rangle 
\nonumber \\
{\hat{C}}_t| \bar{y} \rangle & = &
\langle\bar{y}|{\hat{C}}_t| \bar{y} \rangle | \bar{y} \rangle + 
\sum_{k=0}^{n-1} 
\langle {\bar{y}}_{1k} | {\hat{C}}_t| \bar{y} \rangle | {\bar{y}}_{1k} \rangle  
\nonumber \\
{\hat{C}}_t \hat{W} | \bar{y} \rangle & = &
\langle\bar{y}| \hat{W} {\hat{C}}_t \hat{W} | \bar{y} \rangle \hat{W} | \bar{y} \rangle + 
\sum_{k=0}^{n-1} 
\langle{\bar{y}}_{1k}| \hat{W} {\hat{C}}_t  
\hat{W} | \bar{y} \rangle \hat{W} | {\bar{y}}_{1k} \rangle .
\label{AyAWyCyCWy}
\end{eqnarray}
Here $\{| {\bar{y}}_{1k} \rangle  {\}}_{0 \leq k \leq n-1}$ are the $n$ basis vectors, that are 
different from the basis vector $| \bar{y} \rangle$ in a single qubit only, denoted as 
the $k$'th qubit in the register.  For example, in a 5 qubit system, where 
$| \bar{y} \rangle = | \bar{17} \rangle = |10001\rangle$, 
$\{|{\bar{y}}_{1k} \rangle{\}}_{0 \leq k \leq 4}$  
is the set of computational basis states which includes $|10000\rangle$, $|10011\rangle$,
$|10101\rangle$, $|11001\rangle$ and $|00000\rangle$.
Any operation of either ${\hat{A}}_t$ or    
${\hat{C}}_t$ on basis state vectors of the form  $| \bar{y} \rangle$ or 
$\hat{W}| \bar{y} \rangle$, 
increases the dimension of the relevant vectors space by $n$ additional independent 
directions determined by $\{| {\bar{y}}_{1k} \rangle {\}}_{0 \leq k \leq n-1}$ and 
$\{ \hat{W}| {\bar{y}}_{1k} \rangle  {\}}_{0 \leq k \leq n-1}$ respectively. 

Grover's search algorithm is performed by executing the  
operator $\hat{G} (T)$ on the initial state of zeros $| \bar{0} \rangle$.  Ignoring the effect of the 
unitary noise in the Hadamard transforms (i.e. considering 
$\hat{G} (T) \approx  {\hat {G}}_0 (T) =  \hat{W} {\hat{Q}}^T$ ), one finds that 
Grover's output state vector after time $T$ lays in a 2-dimensional sub-space, 
spanned by the Hadamard operation on the initial state of zeros 
$\hat{W}| \bar{0} \rangle$ and the marked state $| \bar{m} \rangle $  (Note that ${\hat{Q}}^T$ 
rotates the state vector $| \bar{0} \rangle $ in the 2-dimensional sub-space 
spanned by $| \bar{0} \rangle$ and $ \hat{W}| \bar{m} \rangle$ while ${\hat{W}}^2 = \hat{I}$ is 
the identity operator).

The unitary noise in the Hadamard transforms add perturbation elements to 
Grover's operator $\hat{G}_0 (T)$ [see Eq.  (\ref{hat_GT_approx})].  
These elements consist of the noise generators ${\hat{A}}_t$ and ${\hat{C}}_t$, powers of 
the Grover iteration $\hat{Q}$ and selective inversion ${\hat{I}}_0$. 
Executions of $\hat{Q}$ and ${\hat{I}}_0$ do not remain confined to    
a certain 3-dimensional sub-space. However, any operation of 
either ${\hat{A}}_t$ or  ${\hat{C}}_t$ on a basis state vector of the form 
$| \bar{y} \rangle$ or $\hat{W}| \bar{y} \rangle $, 
($| \bar{y} \rangle  = | y_{n-1},y_{n-2}, \ldots y_0 \rangle $ 
where $y_k = 0,1$ ), extends the superposition  by $n$ additional independent state  
vectors.  The higher the order $l$ of the perturbation component ${\hat{G}}_l (T)$ 
[see Eq. (\ref{hat_GT})], the larger the number of executions of the noise generators.  

Consider a realization of the noise in which all the $\alpha$'s and the $\gamma$'s are 
of order $\epsilon$. Table \ref{tb:tab1} shows the vectors which appear in 
the superpositions produced by the leading perturbation components of Grover's 
expansion (\ref{hat_GT_approx}) with their corresponding order of the noise. 
An optimal time $T = T_0$, in which the noiseless Grover's 
operator returns exactly the marked state $| \bar{m} \rangle$ is assumed
(i.e. $\hat{G}_0 (T_0) | \bar{0} \rangle = | \bar{m} \rangle$). 
The vectors  $\{ | {\bar{m}}_{1k} \rangle  {\}}_{0 \leq k \leq n-1}$ and 
$\{ | {\bar{0}}_{1k} \rangle  {\}}_{0 \leq k \leq n-1}$ are the $n$ vectors that are respectively 
different from the marked state $| \bar{m} \rangle$ and the state of zeros $| \bar{0} \rangle$ 
in their $k$'th  bit only, the vectors  
$\{ | {\bar{m}}_{2k_{1}k_{2}} \rangle  {\}}_{0 \leq k_1 < k_2  \leq n-1}$
and  $\{ | {\bar{0}}_{2k_{1}k_{2}} \rangle  {\}}_{0 \leq k_1 < k_2  \leq n-1}$ denote the $C_{2}^{n}$ 
vectors that are respectively different from those vectors in the $k_1$'th and $k_2$'th bits  and so on 
( $C_{l}^{n}$ is the binomial coefficient $C_{l}^{n} = \frac{n!}{l!(n-l)!}$). 

Given the marked state 
$| \bar{m} \rangle$ we define a neighborhood class $l$ which includes all computational 
basis states $| \bar{x} \rangle$ that are different from the marked state $| \bar{m} \rangle$ 
in exactly $l$ bits.   e.g. consider a 5 qubit system with marked state 
$| \bar{m} \rangle = | \bar{17} \rangle = |10001\rangle$, one finds the state  
$| \bar{x} \rangle = | \bar{31} \rangle = |11111\rangle$ in the $3$'rd neighborhood class. 

Table {\ref{tb:tab1} clearly shows that in spite of $T$ being the optimal time $T_0$, the noise causes 
a flow of probability from the marked state $| \bar{m} \rangle$. (In case that $T$ is not 
optimal, the probability flows out of the original 2-dimensional Grover's space, spanned by 
$| \bar{m} \rangle$ and $\hat{W}| \bar{0} \rangle$.) 

A certain portion of the probability 
``diffuses'' from the marked state $| \bar{m} \rangle$ to its  neighbors  
$\{| {\bar{m}}_{1k} \rangle  {\}}_{0 \leq k \leq n-1}$,  
$\{ | {\bar{m}}_{2k_{1}k_{2}} \rangle  {\}}_{0 \leq k_1 < k_2  \leq n-1}$ etc., 
where any perturbation component of order $l$ ($\hat{G}_l (T)$) contributes amplitudes 
of order $O(\epsilon^l)$ to all the states within neighborhood classes of order $l' \leq l$.

Another portion of probability involves the Hadamard operator and hence flows  uniformly to all $N$ 
computational basis states. 
Since the first component which includes a Hadamard operator is $O({\epsilon}^1)$, the contribution 
of this flow to each computational basis state is $O(\frac{\epsilon}{\sqrt{N}})$.

Clearly in the limit of large $N$, the "diffusive" flow into states of small $l$ 
neighborhood class (of order $O({\epsilon}^l)$) is much stronger than the uniform flow 
(of order  $O(\frac{\epsilon}{\sqrt{N}})$). 
We call these states $| \bar{m} \rangle$'s {\it near}  states. All other states are considered 
{\it far} states. 
There is a trade-off; on one hand the probability to measure a certain near state is larger 
than the probability to measure a certain far one. On the other hand there are much more 
far states ($O(N)$ far states where $N=2^n$) than near states  ($O(n^l)$ near states, 
where $l$ is small).

We will now quantify the performance of Grover's search 
algorithm in the presence of unitary noise.  To this end we will calculate 
the average probabilities to measure certain basis states  
$| \bar{x} \rangle =| x_{n-1},x_{n-2}, \ldots x_0 \rangle$ at the optimal measurement time. 
Without noise, at the optimal measurement time the probability to measure the marked state 
is $p_0 = 1$. Thus, the probability of any other state is zero.  According to 
Eq. (\ref{ket_gT}) using the matrix representation in Eq. (\ref{Q^tapprox}), the unperturbed  
probability of measuring an unmarked basis state $| \bar{x} \rangle$ 
(i.e. $| \bar{x} \rangle  \neq | \bar{m} \rangle$) at a certain time $T$ is: 
\begin{equation}
p_0(\bar{x},T) = {|\langle\bar{x}|\hat{G}_{0}(T)| \bar{0} \rangle|}^2 
= {| \frac{1}{\sqrt{N}} \cos(\omega T) + O(\frac{1}{N}) |}^2
\label{p0xT}
\end{equation}
where $N = 2^n$ is the total number of computational basis states. Therefore 
the optimal measurement time is given by:  
\begin{equation}
T_0 = \frac{\pi}{2 \omega} + \frac{O(\frac{1}{\sqrt{N}})}{\omega} 
= \frac{\pi}{4} \sqrt{N} + O(1).
\label{T0}
\end{equation}

Adding a perturbation to Grover's quantum search operator, a second order expansion 
of the probability to measure a state $| \bar{x} \rangle$ at time $T$ gives:
\begin{equation}
p(\bar{x},T) = {|\langle\bar{x}|\hat{G}(T)| \bar{0} \rangle|}^2 
= {|\langle\bar{x}|\hat{G}_{0}(T)| \bar{0} \rangle |}^2 + \Delta p(\bar{x},T) 
\label{pxT}
\end{equation}
where:
\begin{equation}
\Delta p(\bar{x},T) = 
R(\bar{x},T)+ {|\langle\bar{x}|\hat{G}_{1}(T)| \bar{0} \rangle|}^2 
\label{Deltap}
\end{equation}
and
\begin{equation}
R(\bar{x},T) = 
2\langle\bar{x}|\hat{G}_{0}(T)| \bar{0} \rangle
Re\{\langle\bar{x}|\hat{G}_{1}(T)| \bar{0} \rangle+
\langle\bar{x}|\hat{G}_{2}(T)| \bar{0} \rangle\}.
\label{Rp}
\end{equation}
The normalization condition dictates that the probability to measure the marked state 
$| \bar{m} \rangle$ at any time $T$ satisfies
$p(\bar{m},T) = 1 - \sum_{\bar{x} \neq \bar{m}} p(\bar{x},T)$.
When $T$ is around the optimal time $T_0$, the probability to measure 
the marked state approaches unity. 
Therefore, all the terms that are much less than $O(\frac{1}{N})$ and appear in 
the $N-1$ probabilities $p(\bar{x},T)$ to measure unmarked states $| \bar{x} \rangle$ 
respectively,  become negligible. 

For time $T$ close to $T_0$ an expression 
of the form $T = T_0 + t$ where $|t| = O(1)  \ll T_0 = O(\sqrt{N})$ 
can be written.  Hence, for any unmarked state the estimation:
\begin{equation}
|\langle\bar{x}|\hat{G}_{0}(T)| \bar{0} \rangle| = \frac{1}{\sqrt{N}} |\cos(\omega T)|
= \frac{1}{\sqrt{N}} |\sin(\omega t)|= \frac{\omega}{\sqrt{N}} |t|  
= O\left(\frac{1}{N}\right) 
\label{<G0x0>}
\end{equation}
can be made. Therefore, 
${|\langle\bar{x}|\hat{G}_{0} (T) | \bar{0} \rangle|}^2=O(\frac{1}{N^2}) \ll O(\frac{1}{N})$ 
is negligible. 

On the other hand, a small perturbation around 
the original Grover operator $\hat{G}_0 (T)$ is assumed.  Namely,  
$|\langle\bar{x}|\hat{G}_{2}(T)| \bar{0} \rangle| \ll 
|\langle\bar{x}|\hat{G}_{1}(T)| \bar{0}\rangle| \ll O(1)$,  
so that  $|R(\bar{x},T)|  \ll  O\left(\frac{1}{N}\right)$ and is therefore  negligible. 

Hence, the mean probability of measuring a computational basis state  
$|\bar{x}\rangle$ at time $T$ around  
the optimal time $T_0$ is given by:
\begin{equation}
\langle p(\bar{x},T) \rangle  \approx  \left\{ \begin{array}{ll} 
\langle {|\langle\bar{x}|\hat{G}_{1}(T)|\bar{0}\rangle|}^2 \rangle
& \mbox{if  $\bar{x} \neq \bar{m}$}, \\ 
1 - \sum_{\bar{x} \neq \bar{m}} 
 \langle ({|\langle\bar{x}|\hat{G}_{1}(T)|\bar{0}\rangle|}^2) \rangle & \mbox{if $\bar{x} = \bar{m}$} 
\end{array}  \right.
\label{pxT_total}
\end{equation}
where $|\bar{m}\rangle$ is the marked state, $\langle \rangle$ denotes the  averaging on the noise  and
\begin{equation}
\langle\bar{x}|\hat{G}_{1}(T)|\bar{0}\rangle  = 
i \{ \sum_{t=0}^{T} \langle\bar{x}| \hat{W}{\hat{Q}}^{T-t}
({\hat{A}}_{t+1} +{\hat{I}}_0 {\hat{C}}_t {\hat{I}}_0 ) 
{\hat{Q}}^t |\bar{0}\rangle \} 
\label{G1xT}
\end{equation}
using the definition of ${\hat{C}}_{t=0} \equiv 0$.

In the Appendix, we apply the matrix representations which 
appear in Eqs. (\ref{Q^tapprox}) and (\ref{I0approx}) as well as the noise 
matrix elements in Eqs. (\ref{<Axy>})-(\ref{<Cxy>}) and (\ref{<WAxyW>})-(\ref{<WCxyW>})
to calculate the leading terms of the mean measurement probabilities at the optimal measurement 
time in the limit of weak noise and large $N$. We assume that the two Hadamard transforms, 
which appear in each Grover iteration, are implemented with similar hardware. 
We also assume that the noise is unbiased, or alternately we refer to the case in which the exact 
values of the bias elements are known (a-priory) and can be shifted away in every iteration of 
the Hadamard gate. For further simplification we only consider an isotropic noise.   
Thus, statistical moments of the noise defined in 
Eqs.  (\ref{<alpha_gamma>}) and (\ref{<alpha2_gamma2>}) 
takes the form:
\begin{equation}
 \langle \alpha_{\mu} \rangle =  \langle \gamma_{\mu} \rangle = 0
\label{isotropic_1st_condition}
\end{equation}
and 
\begin{equation}
 \langle \delta \alpha_{\mu} \delta \alpha_{\nu} \rangle  = 
 \langle \delta \gamma_{\mu} \delta \gamma_{\nu} \rangle  = \delta_{\mu \nu} {\epsilon}^2,
\label{isotropic_2nd_condition}
\end{equation}
where $\delta_{\mu \nu}$ is the Kronecker's delta function ($\mu,\nu = 1,2,3$) and 
$\epsilon$ is the isotropic noise's standard deviation. It is shown in the Appendix 
(Eqs. (\ref{P0_A})-(\ref{Pfar_A})) that for {\it any} given distribution 
of the noise,  the averaged probabilities at optimal measurement time are approximated 
in the limit of large $N$ and small $\epsilon$ by:
\begin{equation}
P_0 = 1 - \frac{9}{8} \pi n \sqrt{N} {\epsilon}^2 + O(n^2 N {\epsilon}^4)
\label{P0}
\end{equation}
\begin{equation}
P_1 =  \frac{\pi}{2} n \sqrt{N} {\epsilon}^2  + O(n^2 N {\epsilon}^4) 
\label{P1}
\end{equation}
\begin{equation}
P_{far} =  \frac{5}{8} \pi n \sqrt{N} {\epsilon}^2  + O(n^2 N {\epsilon}^4).
\label{Pfar}
\end{equation}
$P_0$ is the mean measurement probability of the marked state $| \bar{m} \rangle$. 
$P_1$ is the averaged probability to measure a {\it near} computational basis state 
which lays in the first marked state's neighborhood class 
(and is hence different from the marked state in a single bit only). 
$P_{far}$ is the averaged probability to measure a {\it far} computational 
basis state that differs from the marked state in more than one bit. 

This result reveals some flexibility which enables us to find the marked state even when 
a searching error has occurred.  In case of weak noise (i.e. if $\epsilon$ is small), 
the mean probabilities to measure near and far states are of the same order. With 
an approximate probability of $\frac{P_1}{P_1 + P_{far}} \approx \frac{4}{9}$, 
a searching error yields a near state which is different from the marked state in a single 
bit only.  By flipping the bits of the measured near state one at a time, the marked state 
can be reconstructed after at most $n$ steps. Similar behavior also exists in the non-isotropic 
case, although the mean probabilities no longer have a compact form. 

A  re-scaled  standard deviation  magnitude:
\begin{equation}
\eta = \sqrt{n \sqrt{N}} \epsilon 
\label{eta}
\end{equation}
may be considered as the parameter of the problem. 
The limit of $\eta \ll 1$ ($\epsilon \ll \frac{1}{\sqrt{n \sqrt{N}}}$) implies for 
{\it weak} noise in which the above 
approximations are valid. The limit of $\eta \gg 1$  ($\epsilon \gg \frac{1}{\sqrt{n \sqrt{N}}}$) 
is the {\it strong} noise limit where the noise completely destroys the quantum search. 
In this case the measured state 
is randomly taken from the total number of $N$ possible measured states, so that 
$P_0 \rightarrow \frac{1}{N}$ and $P_1  \rightarrow \frac{n}{N}$. In between those extremes
we refer to the noise as {\it moderate}.

In the next section we present numerical results which support the analytical predictions. 
We evaluate the performance of the unbiased and isotropic noisy quantum search also in case of 
moderate noise, where the limit of weak noise is not valid anymore.  

\section{The performance of the noisy search algorithm - numerical results}

We first show simulation results confirming the predictions of the analytic approximation.
The simulations also enable us to study the effect of larger values of $\epsilon$  
(the standard deviation of the noise). These results give rise to
new search strategies which we discuss in section V.

The simulations were written in C++ and FORTRAN, on i686 machines running Red Hat Linux. 
We applied isotropic unbiased Gaussian unitary noise by transforming the output of a uniform  
random number generator to be Gaussly distributed. The random Gaussian 
variables $\alpha_\nu$ and $\gamma_\mu$ ($\mu = 1,2,3$) are the coefficients of the 
Pauli matrices as defined in Eq. (\ref{hat:sigma}). 
Note that although these coefficient are limited to the range between
$-\pi$ and $\pi$, the standard deviation of the noise satisfies $\epsilon \ll 1$,
thus the corrections to the Gaussian distribution are negligible.
In order for the results to be 
statistically sound they are averaged over sufficient number of runs. 
The statistical error is estimated by comparing 
the results of different runs with identical parameters.  

Fig. \ref{fg:fig1} shows the dependence of $P_{0}$,$P_{1}$  and $P_{far}$
on the standard deviation of the noise $\epsilon$, for relatively
small values of $\epsilon$.  The predictions of Eqs. (\ref{P0}), (\ref{P1}) and (\ref{Pfar}),
are compared to the simulated results.
The prediction is valid for small values of $\epsilon$ 
($\epsilon <0.006\simeq 0.166\sqrt{\frac{1}{n\sqrt{N}}}$,
for number of qubits in the register $n=12$). In the presence of noise with larger standard deviation, 
the next terms in the approximation are no longer negligible,
and the simulated results start to deviate from the predicted ones.

Fig. \ref{fg:fig2} shows the same dependence for larger values of $\epsilon$.
Results are plotted for system sizes of 8 and 15 bits. Note that a re-scaled 
standard deviation $\eta = \sqrt{n \sqrt{N}} \epsilon$ axis is used as 
predicted in Eq. (\ref{eta}). Indeed, this scaling makes the two graphs diverge.
This further confirms the validity of the approximation. 

We can roughly divide the noise deviation level into three regions:
\begin{itemize}
\item Weak noise deviations: $\eta  \alt0.166$
($\epsilon \alt0.166\sqrt{\frac{1}{n\sqrt{N}}}$).
This level of noise was previously discussed. 
\item Strong noise deviations:  $\eta  \agt 1.4$
($\epsilon \agt 1.4\sqrt{\frac{1}{n\sqrt{N}}}$).
At this level of the noise Grover's algorithm should
completely fail and the probability distribution among the basis states
should be uniform. This means that for large $\epsilon$,
$ P_{0}\rightarrow \frac{1}{N}\, ,\, and\, P_{1}\rightarrow \frac{n}{N}$ which
is verified by the simulation. 
\item Moderate noise deviation: 
$0.166 \alt\eta \alt1.4$ 
($ 0.166\sqrt{\frac{1}{n\sqrt{N}}} \alt \epsilon \alt 1.4\sqrt{\frac{1}{n\sqrt{N}}}$).
The exact behavior of Grover's algorithm under this level of noise 
and the specific boundaries of the moderate noise region are  not
predicted by the analytic approximation. However, the simulation enables
us to study this case as well. As the noise increases, probability
``diffuses'' from the marked state to the near states and 
$P_{1}$ increases. Gradually, the standard deviation of the noise becomes too
large, which makes the probability ``diffuse'' to all states,
near and far, uniformly. In this case the vast number ($O(N)$)
of far states overcomes the few ($n$) near states, and $P_{far}$
approaches one. 
\end{itemize}
Note that there is no noise level for which $ P_{0} $ is negligible
while $ P_{1} $ is not. This is a results of the effect of high
order terms in the approximation. Recall equations (\ref{P0}) and
(\ref{P1}). 
For $ P_{1} $ to be of the order of 1 we need $ \epsilon ^{2}=O\left( \frac{1}{n\sqrt{N}}\right)  $,
and in this case the higher order terms are $ O(1) $ as well, which
implies that $ P_{0} $ is not necessarily negligible.

To demonstrate the flow of probability from the marked
state to near and far states, we divide the computational basis vectors into classes
of neighborhood to the marked state. Given the marked state vector $ \left| \bar{m}\right\rangle  $
we have defined its $l$'th order neighborhood  class as the set of $ C^{n}_{l}=\frac{n!}{l!(n-l)!} $ 
state vectors that differ from $ \left| \bar{m}\right\rangle  $
in exactly $l$ bits. Let $ P_{l}$, $l=0,\ldots ,n $ denote
the probability to measure any of the state vectors of the $l$'th
neighborhood class. Fig. \ref{fg:fig3} shows the distribution of probability among
the $ P_{l} $'s for different values of $\eta$. For
small $\eta$, the marked state (neighborhood class of order
$ 0 $) should be measured with probability 1. The remaining neighborhood
classes should have zero measurement probability. For large $\eta$,
the distribution of probability should be uniform among basis state
vectors, which should reflect in a binomial distribution of probability
among the neighborhood classes (proportional to their size). 
The transformation between the two extremes is determined by the two probability 
flow types caused by the noise: the "diffusive" and the uniform. 
First $ P_{1} $ increases due to the "diffusive" part of the flow. 
Then, the middle (around $ \frac{n}{2} $) neighborhood classes become
noticeable due to the number of elements in these classes.
Yet, there are cases ($ 0.803\leq \eta \leq 1.069 $) where
$ P_{0} $ is quite small, but $ P_{1} $ and $ P_{2} $ are
still relatively large. For even larger values of $\eta$
we have a semi-binomial distribution distorted toward the low order neighborhood
classes, which eventually (for strong noise) becomes an almost pure binomial
distribution.

\section{Alternative search strategies}

One can identify two search strategies using classical or quantum search:

\begin{itemize}
\item Classical search until marked element is found. This strategy has an average
run time of\begin{equation}
\label{eq:Tcalssical}
T_{classical}=\frac{N}{2}
\end{equation}
steps.
\item Quantum search followed by a single classical verification step until
marked element is found. This is a geometric procedure with success
probability $ P_{0} $ and has an average run time of\begin{equation}
\label{eq:Tgrover}
T_{Grover}=\frac{\frac{\pi }{4}\sqrt{N}\tau _{q}+1}{P_{0}}
\end{equation}
where a single classical computation step is performed in one time unit
and a single quantum (Grover) iteration is performed in $ \tau _{q} $ time
units. 
\end{itemize}
The property of ``diffusion'' of probability from the marked state
to the far states through the near ones raises the possibility of
using hybrid (quantum and classical) search strategies. These strategies are  based on classically 
searching the marked element starting from the state measured after applying Grover's
algorithm, and going over the states according to the class of neighborhood
to which they belong. The general hybrid strategy has a parameter
$l$, and is defined as follows:

\begin{enumerate}
\item Run Grover's algorithm, and measure the state of the register  in the computational basis at the optimal measurement time ($\frac{\pi}{4}\sqrt{N}\tau_q$). 
\item Repeat for $ j=0\ldots l $: Classically search the marked state
among the states of the $j$'th-order neighborhood class until it
is found.
\item If the marked element was not found, go back to step 1.
\end{enumerate}
Naturally, the average number of quantum and classical operations required
in order to find the marked element depends on the noise (which controls
the probabilities $ P_{l} $), and on the value of the parameter
$l$. The effectiveness of each strategy is also governed by the
time it takes to perform a single quantum computation step, $ \tau _{q} $
(which is the time required for a single Grover iteration). 

In order to analyze the performance of each strategy we define $ \left\langle T_{l}\right\rangle  $
to be the average time required to complete an entire search of the
marked element using the hybrid strategy with parameter $l$.
We also define $ \Pi _{l} $ to be the probability that the marked
element is found in a search of the first $l$ neighborhood classes.
Obviously, 
\begin{equation}
\Pi _{l}=\sum_{j=0}^l P_{j}
\label{Pil}
\end{equation}
Let $ T_{l} $ denote the average time required to complete a single
execution of steps 1 and 2 above in case that the marked element is
found, and $ \widetilde{T_{l}} $ denote the time required to perform
a single execution of these steps in case the mark element is not
found. These are given by:
\begin{equation}
\widetilde{T_{l}}=\frac{\pi}{4} \sqrt{N} \tau _{q}+\sum ^{l}_{j=0}C_{l}^{n}
\label{Tl_tilde}
\end{equation}
where $ C^{n}_{l} $ is the binomial coefficient $ C_{l}^{n}=\frac{n!}{l!(n-l)!} $,
and
\begin{eqnarray}
T_{l} & = & \frac{1}{\Pi _{l}}\left[ P_{0}(\frac{\pi}{4} \sqrt{N}\tau _{q}+1)+P_{1}(
\frac{\pi}{4} \sqrt{N}\tau _{q}+1+\frac{1}{2}C^{n}_{1})+P_{2}(
\frac{\pi}{4} \sqrt{N}\tau _{q}+1+C^{n}_{1}+\frac{1}{2}C^{n}_{2})+\ldots \right]  \nonumber \\
& = & \frac{\pi}{4} \sqrt{N}\tau _{q}+1+\frac{1}{\Pi _{l}}\sum ^{l}_{j=1}P_{j}
\left( \sum ^{j-1}_{i=1}C^{n}_{i}+\frac{1}{2}C^{n}_{j}\right).
\label{Tk}
\end{eqnarray}
This is the calculation of the mean time, normalized by $ \Pi _{l} $.

We can now express $ \left\langle T_{l}\right\rangle  $ in terms
of $ T_{l}$,$\widetilde{T_{l}} $ and $ \Pi _{l} $:
\begin{eqnarray}
\left\langle T_{l}\right\rangle & = & \Pi _{l}T_{l}+\Pi _{l}(1-\Pi _{l})(T_{l}
+\widetilde{T_{l}})+\Pi _{l}(1-\Pi _{l})^{2}(T_{l}+2\widetilde{T_{l}})+\ldots  \nonumber \\
& = &  \Pi _{l}T_{l}\sum ^{\infty }_{r=0}(1-\Pi _{l})^{r}+\Pi _{l}\widetilde{T_{l}}
\sum ^{\infty }_{r=0}r(1-\Pi _{l})^{r} \nonumber \\
& = & T_{l}+\frac{1-\Pi _{l}}{\Pi _{l}}\widetilde{T_{l}}.
\label{<Tl>}
\end{eqnarray}
The average time of the optimal strategy is given by
\begin{equation}
\left\langle T\right\rangle =\min _{l}\left( \left\langle T_{l}\right\rangle \right). 
\label{<T>}
\end{equation}

Of course we must have $ \left\langle T\right\rangle <\frac{N}{2} $.
Otherwise the quantum search is completely inefficient, and the regular
classical search yields better results. The best strategy parameter
$ l_{opt} $ is:
\begin{equation}
l_{opt}=\arg \left( \min _{l} \left\langle T_{l}\right\rangle \right), 
\label{k_opt}
\end{equation}
where $\arg$ denotes the argument of the minimum function, i.e.
the value of $l$ for which 
$\langle T_{l} \rangle$
is minimal.

We work under the (strict) assumption that $ \tau _{q}=1 $, which is
that a quantum computation step requires the same amount of time
as a classical computation step. Fig. \ref{fg:fig4} shows the optimal  
strategy averaged time  $ \left\langle T\right\rangle  $,
calculated from the data presented in Fig. \ref{fg:fig3}, as a function of the
noise's re-scaled standard deviation $\eta = \sqrt{n {\sqrt{N}}}\epsilon$ 
It also shows the performance of the two
trivial strategies mentioned above (Eqs. \ref{eq:Tcalssical} and \ref{eq:Tgrover}).
It is apparent that the required time is decreased by up to a factor
of $ 7 $. It is especially important to note that for noise deviation 
$ \eta \agt 1.4 $ ($\epsilon \agt 1.4 \frac{1}{\sqrt{n \sqrt{N}}}$)
the original Grover quantum search is useless. 
(At the point $\eta \simeq 1.4$ the curve of the naive quantum searching time 
intersects with the $\frac{1}{2} N$ line which is the averaged time of a classical 
search).
On the other hand, the optimal hybrid
strategy can be used to substantially decrease the required time up to 
$ \eta \simeq 1.9 $ ($\epsilon \simeq 1.9 \frac{1}{\sqrt{n \sqrt{N}}}$).
It is obvious that under large noise deviations the optimal hybrid strategy 
requires $\frac{\pi \sqrt{N}}{4}\tau_{q}+\frac{N}{2}\simeq\frac{N}{2} $ time to complete the search (this is the time required for an initial quantum search
followed by a classical search of the entire search domain). 
As it approaches this limit ( $ \eta \agt 1.9 $), the dependency of 
the time required by the hybrid strategy becomes a concave function 
of $ \eta $.
Hence, it is for this noise levels only ( $ 1.4 \alt\eta \alt1.9 $ ) that the
hybrid strategy yields significant improvements. Above this level of
noise, the classical search is at least as efficient (and does not require the initial
quantum search).

Table \ref{tb:tab2} shows the chosen strategy $ l_{opt} $ calculated for 
each level of
noise. It can be seen that as the noise's standard deviation increases,
the chosen $ l $ increases as well. This means that it is worthwhile
to classically search increasingly more distant neighborhood classes for 
the marked element. For large  $ \eta $, $ l_{opt}=20 $, which means 
that the chosen strategy is not better than the naive classical search.

This analysis shows that the property of ``diffusion'' of probability
from the marked state to the near states can be used to obtain a hybrid
search strategy which is more efficient than the naive quantum search
strategy. Note that the results above were calculated for $ \tau _{q}=1 $.
If we make a more reasonable assumption $ \tau _{q}\gg 1 $ (i.e.,
that a single quantum computation step takes more time than a classical
one), the improvement would be considerably better. In addition, the
hybrid strategy enables an efficient quantum search under noise levels
for which the naive quantum strategy fails.

\section{Summary}
We have studied the effect of unitary noise on the performance of Grover's quantum search 
algorithm. This type of noise may result from tiny fluctuations and drift in
the parameters of the (quantum) components performing the computation.
The resulting operations are still unitary, but not precisely those
assumed in the design of the algorithm. 
In the analysis we focused on the effect
of an unbiased unitary noise in the Hadamard gate $W$.
For simplicity we further assumed that the noise is isotropic as well.
The Hadamard gate is an essential component in each iteration of the
quantum search process. 
The gate $W$ was replaced by a noisy Hadamard
gate $U$, whose parameters are distributed around those of $W$
according to an unbiased, symmetric probability distribution. 
The noise level was characterized by its standard deviation.  

It was found that for noise levels greater than 
$O(n^{-\frac{1}{2}} N^{-\frac{1}{4}})$, Grover's algorithm becomes 
inefficient. (Here $n$ denotes the number of qubits in the register and $N = 2^n$ 
is the number of computational basis states). The nature of the flow of probability 
out of the marked state, which is caused by the noise 
was also investigated, analytically and numerically. A phenomenon of "diffusive" 
flow of probability to the marked state's near neighbors, that are different from 
the marked state in only a few bits, was observed. This feature of ``diffusion'' 
gives rise to new hybrid search strategies which are shown to improve the 
effectiveness of the entire search procedure, both in terms of the time required 
to complete the search, and in the noise level under 
which a quantum search is still more effective than the classical search.
The use of hybrid strategies was found efficient even under noise levels 
for which naive re-execution of Grover's quantum search (until the marked state is found) 
fails. 

\section{Acknowledgments}
This work was supported by EU fifth framework program Grant No. IST-1999-11234. 

\appendix  
\section{Calculation of mean probabilities}
We are aimed to obtain an expression for the averaged probabilities 
to measure a state $|\bar{x}\rangle$ around the optimal measurement 
time in the presence of one qubit unitary noise in the Hadamard 
gate.  We begin with Eq. (\ref{G1xT}). As one can see 
$\langle\bar{x}|\hat{G}_{1}(T)|\bar{0}\rangle$ is built as sum over time index 
$t$ of matrix elements produced by 4 consequent operations on the initial 
state of zeros $|\bar{0}\rangle$:  ${\hat{Q}}^t$,
$({\hat{A}}_{t+1} +{\hat{I}}_0 {\hat{C}}_t {\hat{I}}_0 )$, ${\hat{Q}}^{T-t}$ 
and $\hat{W}$ in that order. Using the matrix representations of  ${\hat{Q}}^t$ 
and ${\hat{I}}_0$ in Eqs. (\ref{Q^tapprox}) and (\ref{I0approx}), as well as the 
Hermitian noise generators presented in Eq. (\ref{AyAWyCyCWy}), the following 
operations can be expanded: 
\begin{equation}
{\hat{Q}}^t |\bar{0}\rangle = \cos(\omega t) |\bar{0}\rangle 
+ \sin(\omega t) \hat{W}|\bar{m}\rangle + \vec{O}
\label{Q^t|0>}
\end{equation}
\begin{equation}
({\hat{A}}_{t+1} +{\hat{I}}_0 {\hat{C}}_t {\hat{I}}_0 )|\bar{0}\rangle = 
\langle\bar{0}|({\hat{A}}_{t+1} + {\hat{C}}_t ) |\bar{0}\rangle|\bar{0}\rangle + 
\sum_{k=0}^{n-1}   
\langle{\bar{0}}_{1k}|(\hat{A}_{t+1}-\hat{C}_t)|\bar{0}\rangle
|{\bar{0}}_{1k}\rangle
\label{A+I0CI0|0>}
\end{equation}
\begin{eqnarray}
({\hat{A}}_{t+1} +{\hat{I}}_0 {\hat{C}}_t {\hat{I}}_0)  \hat{W}|\bar{m}\rangle  & = & 
\langle\bar{m}|\hat{W}(\hat{A}_{t+1}+\hat{C}_t)\hat{W}
|\bar{m}\rangle \hat{W}|\bar{m}\rangle  \nonumber \\ 
& & + \sum_{k=0}^{n-1} \langle{\bar{m}}_{1k}|\hat{W}(\hat{A}_{t+1}+\hat{C}_t)\hat{W}
|\bar{m}\rangle \hat{W}|{\bar{m}}_{1k}\rangle   + {\vec{O}}_0
\label{A+I0CI0Wm>}
\end{eqnarray}
\begin{equation}
{\hat{Q}}^{T-t} |\bar{0}\rangle = \cos(\omega [T-t]) |\bar{0}\rangle +
\sin(\omega [T-t])  \hat{W}|\bar{m}\rangle + \vec{O}
\label{Q^T-t|0>}
\end{equation}
\begin{equation}
{\hat{Q}}^{T-t} |\bar{0}_{1k}\rangle = {(-1)}^{T-t}  |\bar{0}_{1k}\rangle + \vec{O} 
\label{Q^T-t|01>}
\end{equation} 
\begin{equation}
{\hat{Q}}^{T-t} \hat{W} |\bar{m}\rangle =  -\sin(\omega [T-t]) |\bar{0}\rangle +
\cos(\omega [T-t])  \hat{W}|\bar{m}\rangle + \vec{O}
\label{Q^T-tW|m>}
\end{equation}
\begin{equation}
{\hat{Q}}^{T-t} \hat{W}|\bar{m}_{1k}\rangle = 
{(-1)}^{T-t}  \hat{W} |\bar{m}_{1k}\rangle + \vec{O} 
\label{Q^T-tW|m1>}
\end{equation} 
where  ${\bar{0}}_{1k}$  and ${\bar{m}}_{1k}$ are the binary strings 
which are respectively different from the initial string of zeros $\bar{0}$ and 
the marked string $\bar{m}$ only in their  $k$'th bit,  
$\vec{O} = 
O\left(\frac{1}{\sqrt{N}}\right) |\bar{0}\rangle 
+ O\left(\frac{1}{\sqrt{N}}\right) \hat{W} |\bar{m}\rangle$, 
and 
${\vec{O}}_0 = O(\frac{n \epsilon}{\sqrt{N}}) | \bar{0} \rangle$ where $O(\epsilon)$ 
denotes the order of the noise.  
Then, performing consequent substitutions of Eqs. (\ref{Q^t|0>})-(\ref{Q^T-tW|m1>}) 
in Eq. (\ref{G1xT}) while memorizing that 
$\langle\bar{x}|\hat{W}|\bar{0}\rangle = \frac{1}{\sqrt{N}}$ and 
$\langle\bar{x}|\hat{W}|{\bar{0}}_{1k}\rangle = {(-1)}^{x_k}\frac{1}{\sqrt{N}}$
($x_k$ is the binary value of the $k$'th bit in the string $\bar{x} = x_{n-1},x_{n-2}, \ldots ,x_0$) 
one finds that for any unmarked state $|\bar{x}\rangle$:
\begin{eqnarray}
\langle\bar{x}|\hat{G}_{1}(T)|\bar{0}\rangle 
& =  \frac{i}{\sqrt{N}} \{ &
\sum_{t=1}^{T} \cos{(\omega t)}\cos{(\omega[T-t])}
\langle\bar{0}|(\hat{A}_{t+1} + \hat{C}_t)|\bar{0}\rangle \nonumber \\ 
& &-\sum_{t=1}^{T} \sin{(\omega t)}\sin{(\omega[T-t])}
\langle\bar{m}|\hat{W}(\hat{A}_{t+1}+\hat{C}_t)\hat{W}
|\bar{m}\rangle \nonumber \\
& & + \sum_{t=1}^{T} \cos{(\omega t)}{(-1)}^{T-t} 
[\sum_{k=0}^{n-1} {(-1)}^{x_k}
\langle{\bar{0}}_{1k}|(\hat{A}_{t+1}-\hat{C}_t)|\bar{0}\rangle ]\} \nonumber \\ 
& + i & \{ \sum_{k=0}^{n-1} \langle\bar{x}|{\bar{m}}_{1k}\rangle 
\sum_{t=1}^{T} \sin{(\omega t)}{(-1)}^{T-t} 
\langle{\bar{m}}_{1k}|\hat{W}(\hat{A}_{t+1}+\hat{C}_t)\hat{W}
|\bar{m}\rangle \} \nonumber \\ 
&   & +O\left(\frac{T n \epsilon}{N}\right)  \nonumber \\ 
& = & O\left(\frac{T n \epsilon}{\sqrt{N}}\right) +  
\sum_{k=0}^{n-1} \langle\bar{x}|{\bar{m}}_{1k}\rangle 
O\left(T n \epsilon \right),
\label{G1xT_trigoA}
\end{eqnarray}
where according to Eqs. (\ref{<Axy>})-(\ref{<WCxyW>}): 
\begin{equation}
\langle\bar{0}|(\hat{A}_{t+1} + \hat{C}_t)|\bar{0}\rangle = 
\sum_{k=0}^{n-1} 
\{ \alpha_{0k} (t+1) +  \alpha_{3k} (t+1) + \gamma_{0k} (t) +  \gamma_{3k} (t) \} 
\label{<0|A+C|0>}
\end{equation}
\begin{equation}
\langle\bar{m}|\hat{W}(\hat{A}_{t+1} + \hat{C}_t)\hat{W}|\bar{m}\rangle = 
\sum_{k=0}^{n-1} 
\{ \alpha_{0k} (t+1)+{(-1)}^{m_k}\alpha_{3k} (t+1) + 
\gamma_{0k} (t)+ {(-1)}^{m_k} \gamma_{3k} (t) \} 
\label{<m|WA+CW|m>}
\end{equation}
\begin{equation}
\langle{\bar{0}}_{1k}|(\hat{A}_{t+1}-\hat{C}_t)|\bar{0}\rangle = 
 \alpha_{1k} (t+1)+i \alpha_{2k} (t+1)-  \gamma_{1k} (t+1)-i \gamma_{2k} (t+1)
\label{<01|A-C|0>}
\end{equation}
\begin{equation} 
\langle{\bar{m}}_{1k}|\hat{W}(\hat{A}_{t+1}+\hat{C}_t)\hat{W}|\bar{m}\rangle = 
 \alpha_{3k} (t+1)- i{(-1)}^{m_k}\alpha_{2k} (t+1) + 
\gamma_{3k} (t) - i{(-1)}^{m_k} \gamma_{2k} (t).   
\label{<m1|WA+CW|m>}
\end{equation}
We immediately observe that the order of $\langle\bar{x}|\hat{G}_{1}(T)|\bar{0}\rangle$ 
depends on  the Hamming distance of the state $|\bar{x}\rangle$ to the 
marked state $|\bar{m}\rangle$. The order of ``far'' states, which denote binary strings 
different from the marked string in more than one bit is $O(\sqrt{N})$ times smaller 
than the order of ``near'' states whose binary strings are different 
from the marked string in exactly one bit. 

According to Eq.  (\ref{pxT_total}), we focus  on computing the mean value of 
${|\langle \bar{x} | \hat{G}_{1}(T)|\bar{0}\rangle|}^2$ around the optimal measurement 
time. For this purpose we evaluate trigonometric sums which appear in 
$\langle {|\langle\bar{x}| \hat{G}_{1}(T)|\bar{0}\rangle|}^2 \rangle $, assuming that 
$\omega T = \frac{\pi}{2} + O\left(\frac{1}{\sqrt{N}}\right)$ 
where $\omega \approx \frac{2}{\sqrt{N}}$. 

For any small angular frequency $\omega$ and arbitrary phase $\varphi$:
\begin{eqnarray}
\sum_{t=1}^{T} \cos(\omega t + \varphi) {(-1)}^{T-t} & = & 
\frac{\cos(\omega T + \varphi) + \cos(\omega [T+1] + \varphi) 
- {(-1)}^{T} \{\cos(\omega + \varphi) + \cos \varphi \} }
{2 (1 + \cos \omega)} \nonumber \\ 
& = & O(1) 
\label{general_cos_sum}
\end{eqnarray}
Therefore:
\begin{equation}
\sum_{t=1}^{T} \cos(\omega t) {(-1)}^{T-t} = O(1)
\label{-1^tcos}
\end{equation}
 \begin{equation}
\sum_{t=1}^{T} \sin(\omega t) {(-1)}^{T-t} = O(1)
\label{-1^tsin}
\end{equation}
and:
\begin{equation}
\sum_{t=1}^{T} {\cos}^2 (\omega t)  \cos (\omega [T-t]) {(-1)}^{T-t} = O(1)
\label{cos2cos-1}
\end{equation}
\begin{equation}
\sum_{t=1}^{T} \sin(\omega t) \cos(\omega t) \cos(\omega [T-t]) {(-1)}^{T-t} = O(1)
\label{sincoscos-1}
\end{equation}
 \begin{equation}
\sum_{t=1}^{T} \sin(\omega t) \cos(\omega t) \sin(\omega [T-t]) {(-1)}^{T-t} = O(1).
\label{sincossin-1}
\end{equation}
(Note that any multiplication of sines and cosines can be reduced 
to sums of sines and cosines).

On the other hand:
\begin{equation}
\sum_{t=1}^{T} {\sin}^2 (\omega t) = 
\frac{T}{2} - \frac{\sin(2 \omega T)}{4 \tan \omega} + \frac{1}{4}\{1 - \cos(2 \omega T)\}
= \frac{\pi}{8} \sqrt{N} + O(1) 
\label{sin2}
\end{equation}
\begin{equation}
\sum_{t=1}^{T} {\cos}^2 (\omega t) = 
\frac{T}{2} + \frac{\sin(2 \omega T)}{4 \tan \omega} - \frac{1}{4}\{1 - \cos(2 \omega T)\}
= \frac{\pi}{8} \sqrt{N} + O(1) 
\label{cos2}
\end{equation}
\begin{equation}
\sum_{t=1}^{T} \cos(\omega t) \cos(\omega [T-t]) =
\frac{T}{2} \cos(\omega T) +  \frac{\sin(\omega T)}{2 \tan \omega}
= \frac{1}{4} \sqrt{N} +  O(1) 
\label{coscos}
\end{equation}
\begin{equation}
\sum_{t=1}^{T} \sin(\omega t) \sin(\omega [T-t]) =
-\frac{T}{2} \cos(\omega T) +  \frac{\sin(\omega T)}{2 \tan \omega}
= \frac{1}{4} \sqrt{N} +  O(1) 
\label{sinsin}
\end{equation}
\begin{eqnarray}
\sum_{t=1}^{T}  {\cos}^2 (\omega t) {\cos}^2 (\omega [T-t]) 
& = &
\frac{T}{4} \{1+\frac{1}{2}\cos{(2 \omega T)}\} + 
\frac{\sin(2 \omega T)}{4}\{\frac{1}{\tan(2 \omega)}+\frac{1}{\tan \omega}\}  \nonumber \\
& = & 
\frac{\pi}{32} \sqrt{N} +  O(1) 
\label{cos2cos2}
\end{eqnarray}
\begin{eqnarray}
\sum_{t=1}^{T}  {\sin}^2 (\omega t) {\sin}^2 (\omega [T-t]) 
& = &
\frac{T}{4} \{1+\frac{1}{2}\cos{(2 \omega T)}\} + 
\frac{\sin(2 \omega T)}{4}\{\frac{1}{\tan(2 \omega)}-\frac{1}{\tan \omega}\}  \nonumber \\
& = & 
\frac{\pi}{32} \sqrt{N} + O(1)  
\label{sin2sin2}
\end{eqnarray}
\begin{eqnarray}
\sum_{t=1}^{T} \cos(\omega t) \cos(\omega [T-t]) \sin(\omega t) \sin(\omega [T-t]) 
& = & 
-\frac{T}{8} \cos(2 \omega T) + \frac{\sin(2 \omega T)}{8 \tan(2 \omega)} \nonumber \\
& = & 
\frac{\pi}{32} \sqrt{N} + O(1).
\label{coscossinsin}
\end{eqnarray}

We have now completed all the needed expansions for expressing 
the mean probability to measure a certain unmarked state $|\bar{x}\rangle$ 
around the optimal measurement time. By multiplying Eq. (\ref{G1xT_trigoA})
with its complex conjugate and averaging the value of 
 ${|\langle\bar{x}|\hat{G}_{1}(T)|\bar{0}\rangle|}^2$ 
using the first and the second moments of the noise 
(see Eqs. (\ref{<alpha_gamma>})-(\ref{<alpha2_gamma2>}))
while leaving the leading terms only, one obtains that the mean probability to measure 
an unmarked state $| \bar{x} \rangle$ around the optimal measurement time is:
\begin{eqnarray}
p_1 & \approx & \frac{\pi}{8} \sqrt{N} 
\{\langle \delta {\alpha_2}^2 \rangle  + \langle \delta {\alpha_3}^2 \rangle +
\langle \delta {\gamma_2}^2 \rangle + \langle \delta {\gamma_3}^2 \rangle\} +  \nonumber \\
& &  \frac{1}{16} 
{\{n[\langle \alpha_3 \rangle +\langle \gamma_3 \rangle ]-
f(\bar{m})[\langle \alpha_1 \rangle +\langle \gamma_1 \rangle ]\}}^2 + \nonumber \\
& &  (\langle \alpha_3 \rangle +\langle \gamma_3 \rangle)^2 
+  (\langle \alpha_2 \rangle +\langle \gamma_2 \rangle)^2 
\label{p1}
\end{eqnarray}
in case that $| \bar{x} \rangle$ is a first order neighbor of the marked state $| \bar{m} \rangle$ 
(i.e $|\bar{x}\rangle = |{\bar{m}}_{1k}\rangle$, is different from the marked state only in its 
$k$'th bit) and
\begin{eqnarray}
p_{far}& \approx \frac{1}{16} &
{\{n[\langle \alpha_3 \rangle +\langle \gamma_3 \rangle ]-
f(\bar{m})[\langle \alpha_1 \rangle +\langle \gamma_1 \rangle ]\}}^2 + \nonumber \\
& \mbox{ }  \frac{\pi}{32\sqrt{N}} \{ &
n[5(\langle \delta {\alpha_1}^2 \rangle  + \langle \delta {\gamma_1}^2 \rangle) + 
4(\langle \delta {\alpha_2}^2 \rangle  + \langle \delta {\gamma_2}^2 \rangle) + 
(\langle \delta {\alpha_3}^2 \rangle + \langle \delta {\gamma_3}^2 \rangle )] \nonumber \\
& & 
f(\bar{m})[\langle \delta \alpha_1 \delta \alpha_3 \rangle + 
\langle \delta \gamma_1 \delta \gamma_3 \rangle] \} 
\label{p_far}
\end{eqnarray}
in case that $| \bar{x} \rangle$  is far state (i.e. the state $|\bar{x}\rangle$ differs from the marked 
state  $|\bar{m}\rangle$ in more than one bit). 
The $\alpha_\mu$'s and $\gamma_\mu$'s ($\mu = 1,2,3$) denote 
the real stochastic variables taken from any arbitrary distribution which characterize the noise, 
(The moments of $\alpha_0$ and $\gamma_0$ vanish because they act as a global phase). 
$n$ is the number of qubits in the register, 
$f(\bar{m})$ is the difference between the number of zeros 
and the number of ones that appear in the binary representation 
of the marked string $\bar{m}$ and $N = 2^n$.

In the noiseless search algorithm, the $n$-qubit  
Hadamard transform (that consists of $n$ one qubit gates) appears twice 
in each iteration. 
These $n$ operations are expected to be implemented with similar hardware. 
The noise characteristics of all one qubit gates are expected to be similar 
but with no correlations between each other. 
The noisy unitary operators  ${\hat{U}}_t$ and  ${\hat{V}}_t$ at any Grover's 
iteration $t$ can therefore be considered as direct multiplications  
of $n$ one-qubit unitary and stochastic operators of the same 
statistical behavior. Thus, according to Eq.  (\ref{hat_UVt})
the statistical properties of   ${\hat{c}}_k (t)$ 
and $ {\hat{w}}_k{\hat{a}}_k (t) {\hat{w}}_k$ are alike for any 
time index $t$ and qubit index $k$.  Here  ${\hat{w}}_k$ denotes 
a single-qubit operator, acting on the $k$'th qubit i.e.:
\[
{\hat{w}}_k = \frac{1}{\sqrt{2}} ({\hat{\sigma}}_{1k} + {\hat{\sigma}}_{3k})
\]
where ${\hat{\sigma}}_{1k}$ and ${\hat{\sigma}}_{3k}$ are  
Pauli operators which act on the $k$'th qubit.  Using the expansions 
of  ${\hat{a}}_k (t)$ and  ${\hat{c}}_k (t)$ by Pauli operators (see Eq. (\ref{hat_act})) 
with the aid of the identity;
\[
(\vec{v_1} \cdot \hat{\vec{\sigma}})(\vec{v_2} \cdot \hat{\vec{\sigma}})   
= (\vec{v_1} \cdot \vec{v_2}) {\hat{\sigma}}_0 + 
i (\vec{v_1} \times \vec{v_2}) \cdot \hat{\vec{\sigma}} 
\]
where $\vec{v_1}$ and $\vec{v_2}$ are 3-dimensional vectors of real numbers, 
$\hat{\vec{\sigma}} = ({\hat{\sigma}}_1,{\hat{\sigma}}_2,{\hat{\sigma}}_3)$ 
is the vector of Pauli operators, ${\hat{\sigma}}_0$ is the identity operator,  
and $\cdot$ and $\times$ denote scalar and vector products respectively, 
one obtains the following relations between statistical moments:
\begin{equation}
\langle \gamma_1 \rangle  = \langle \alpha_3 \rangle  
\hspace{.5in}
\langle \gamma_2 \rangle  = - \langle \alpha_2 \rangle 
\hspace{.5in}
 \langle \gamma_3 \rangle   =  \langle \alpha_1 \rangle  
\label{first}
\end{equation}
and 
\[
\langle \delta {\gamma_1}^2 \rangle  =  \langle \delta {\alpha_3}^2 \rangle 
\hspace{.5in} 
 \langle \delta {\gamma_2}^2 \rangle  = \langle \delta {\alpha_2}^2 \rangle 
\hspace{.5in} 
\langle \delta {\gamma_3}^2 \rangle  = \langle \delta {\alpha_1}^2 \rangle 
\]
\begin{equation}
\langle \delta\gamma_1 \delta\gamma_2 \rangle  = -\langle \delta\alpha_2 \delta\alpha_3 \rangle 
\hspace{.5in} 
\langle \delta\gamma_1 \delta\gamma_3 \rangle =  \langle \delta\alpha_1 \delta\alpha_3 \rangle 
\hspace{.5in} 
\langle \delta\gamma_2 \delta\gamma_3 \rangle = - \langle \delta\alpha_1 \delta\alpha_2 \rangle. 
\label{second}
\end{equation}

Moreover, further simplification can be done, if we assume that the noise is unbiased, 
or alternately that the exact value of the bias of the noise is known 
and can be shifted in every operation of the Hadamard gate. 
We also assume that the physical system which realizes 
the quantum computer is happens to be isotropic. Thus, the statistical moments of the noise 
become:
\[
 \langle \alpha_{\mu} \rangle =  \langle \gamma_{\mu} \rangle = 0
\label{isotropic_1st_conditiona}
\]
and 
\[
 \langle \delta \alpha_{\mu} \delta \alpha_{\nu} \rangle  = 
 \langle \delta \gamma_{\mu} \delta \gamma_{\nu} \rangle  = \delta_{\mu \nu} {\epsilon}^2,
\label{isotropic_2nd_conditiona}
\]
where $\delta_{\mu \nu}$ is the Kronecker's delta function ($\mu,\nu = 1,2,3$) and 
$\epsilon$ is the isotropic noise's standard deviation. Then, substituting these relations in 
Eqs. (\ref{p1}) and (\ref{p_far}) one gets:
\begin{equation}
p_1 \approx \frac{\pi}{2} \sqrt{N} {\epsilon}^2
\label{p1_iso}
\end{equation}
and 
\begin{equation}
p_{far}  \approx \frac{5}{8} \pi \frac{n}{\sqrt{N}} {\epsilon}^2.
\label{p_far_iso}
\end{equation}
Since there are $n$ first order neighbors to the marked state $| \bar{m} \rangle$ and 
$N-n-1$ far states, we find that:
\begin{equation}
P_1 = n p_1 \approx \frac{\pi}{2} n \sqrt{N} {\epsilon}^2
\label{P1_approx_A}
\end{equation}
is the mean probability to measure a first order marked state's neighbor in the 
optimal measurement time and
\begin{equation}
P_{far} \approx  N p_{far} \approx \frac{5}{8} \pi n \sqrt{N} {\epsilon}^2
\label{Pfar_approx_A}
\end{equation}
is mean measurement probability of an unmarked far state (in the optimal time as well). 
The mean probability to measure the marked state is given by the normalization 
condition:
\begin{equation}
P_0 = 1 - P_1 - P_{far} \approx  1 - \frac{9}{8} \pi n \sqrt{N} {\epsilon}^2.
\label{P0_approx_A}
\end{equation}

In order to evaluate the limits in which the above approximations are still valid, 
we have to estimate the order of the residual component produced by higher terms of the noise. 
In our calculation we have focused on the case of an unbiased noise. Therefore, only even 
powers of the noise standard deviation $\epsilon$ appear in the mean probabilities expansions, 
so the next term in the above approximations is $O({\epsilon}^4)$. By taking those terms 
into account, Eq. (\ref{pxT_total}) has the form:
\begin{eqnarray}
\langle p(\bar{x},T) \rangle  & \approx & 
\langle {|\langle\bar{x}|\hat{G}_{1}(T)|\bar{0}\rangle|}^2 \rangle + 
\langle {|\langle\bar{x}|\hat{G}_{2}(T)|\bar{0}\rangle|}^2 \rangle + \nonumber \\
& & \langle (\langle\bar{x}|\hat{G}_{1}(T)|\bar{0}\rangle
{\langle\bar{x}|\hat{G}_{3}(T)|\bar{0}\rangle}^*) \rangle + 
 \langle ({\langle\bar{x}|\hat{G}_{1}(T)|\bar{0}\rangle}^*
\langle\bar{x}|\hat{G}_{3}(T)|\bar{0}\rangle)\rangle 
\label{pxT_next_terms}
\end{eqnarray}
where $\langle p(\bar{x},T) \rangle$ is the mean probability to measure a certain 
computational basis state at time $T$ around the optimal measurement time, 
and $\hat{G}_{1}(T)$,   $\hat{G}_{2}(T)$ and $\hat{G}_{3}(T)$ are Grover's 
perturbation components of the first, second and third order respectively, as defined in 
Eq. (\ref{hat_GT_approx}). 

An explicit calculation shows that in case that the measured state $| \bar{x}\rangle$ 
is a marked state first order neighbor (i.e. $|\bar{x}\rangle = |{\bar{m}}_{1k}\rangle$ is 
different from the marked state $|\bar{m}\rangle$ in the $k$'th bit only):
\begin{equation}
O(\langle\bar{x}|\hat{G}_{1}(T)|\bar{0}\rangle \langle\bar{x}|\hat{G}_{3}(T)|\bar{0}\rangle) = 
O( {|\langle\bar{x}|\hat{G}_{2}(T)|\bar{0}\rangle|}^2) = O(T^2 n {\epsilon}^4)
\label{G123_neighbor}
\end{equation}
while in case that $| \bar{x}\rangle$ is far from the marked state $|\bar{m}\rangle$:
\begin{equation}
O(\langle\bar{x}|\hat{G}_{1}(T)|\bar{0}\rangle \langle\bar{x}|\hat{G}_{3}(T)|\bar{0}\rangle) = 
O( {|\langle\bar{x}|\hat{G}_{2}(T)|\bar{0}\rangle|}^2) = O(\frac{T^2 n^2 {\epsilon}^4}{N}).
\label{G123_far}
\end{equation} 
Thus, due to the fact that there are $n$ first order neighbors of the marked state and 
$N-n-1 \approx N$ far states, one finds that the mean measurement probabilities at time 
$T$ around the optimal measurement time (such that  $O(T) = O(\sqrt{N})$) are given by:
\begin{equation}
P_0 = 1 - \frac{9}{8} \pi n \sqrt{N} {\epsilon}^2 + O(n^2 N {\epsilon}^4)
\label{P0_A}
\end{equation}
\begin{equation}
P_1 =  \frac{\pi}{2} n \sqrt{N} {\epsilon}^2  + O(n^2 N {\epsilon}^4) 
\label{P1_A}
\end{equation}
\begin{equation}
P_{far} =  \frac{5}{8} \pi n \sqrt{N} {\epsilon}^2  + O(n^2 N {\epsilon}^4)
\label{Pfar_A}
\end{equation} 
where $P_0$, $P_1$ and $P_{far}$ are the mean probabilities to measure the marked state, 
a first order neighbor of the marked state and a far state respectively.


\newpage

\begin{table}[t]
\caption{
The list of vectors which appear in the superpositions produced by the leading 
perturbation components of Grover's expansion  (\ref{hat_GT_approx}) 
with their corresponding order of the noise. 
The first row of the table denotes the noiseless Grover's search result at the optimal time 
$T_0$, which satisfies  by definition $\hat{G}_0 (T_0)|\bar{0}\rangle = |\bar{m}\rangle$. 
The second and third the row lists the vectors which appear in the superpositions produced by 
the first and the second order perturbation components $\hat{G}_1$ and $\hat{G}_2$ respectively.   
}
\begin{center}
\begin{tabular}{|l||l|l|} \hline
Perturbation component & Order of the noise & Vectors in the superposition \\ \hline
$\hat{G}_0(T) |\bar{0}\rangle$ & $O({\epsilon}^0)$ & 
$| \bar{m} \rangle$\footnote{In case that $T$ is not the optimal measurement time 
the vector $\hat{W}|\bar{0}\rangle$ also appears in the list.} \\ \hline
$\hat{G}_1(T) |\bar{0}\rangle$ & $O({\epsilon}^1)$ & 
$| \bar{m} \rangle$,  $\{ | {\bar{m}}_{1k} \rangle  {\}}_{0 \leq k \leq n-1}$ \\
& & $ \hat{W}|\bar{0}\rangle$, $\{ \hat{W}|{\bar{0}}_{1k}\rangle {\}}_{0 \leq k \leq n-1}$ \\ \hline 
$\hat{G}_2(T) |\bar{0}\rangle$ & $O({\epsilon}^2)$ & 
$ | \bar{m} \rangle$,  $\{ | {\bar{m}}_{1k} \rangle  {\}}_{0 \leq k \leq n-1}$, 
$\{ | {\bar{m}}_{2k_{1}k_{2}} \rangle  {\}}_{0 \leq k_1 < k_2  \leq n-1}$ \\
& & $\hat{W}|\bar{0}\rangle$, $\{ \hat{W}|{\bar{0}}_{1k}\rangle {\}}_{0 \leq k \leq n-1}$, 
$\{ \hat{W}|{\bar{0}}_{2k_{1}k_{2}}\rangle{\}}_{0 \leq k_1 < k_2  \leq n-1}$ \\ \hline 
$\vdots$ & $\vdots$ & $\vdots$ \\ \hline 
\end{tabular}
\end{center}
\label{tb:tab1}
\end{table}

\begin{table}[t]
\caption{The optimal strategy $ l_{opt} $ for different levels of noise deviation. The calculations are 
based on the data shown on Fig. \ref{fg:fig3} (number of qubits is 20). 
The noise re-scaled standard deviation is $\eta =  \sqrt{n \sqrt{N}} \epsilon$.}
\begin{center}
\begin{tabular}{|c|c|c|c|c|c|c|c|c|c|} \hline
$ \eta $ & 0.0053 & 0.138 & 0.271 & 0.404 & 0.537 & 0.670 & 0.803 & 0.936 & 1.069 \\ \hline
$ l_{opt} $ & 0 & 1 & 1 & 1 & 1 & 2 & 2 & 2 & 2 \\ \hline \hline
$ \eta $ & 1.202 & 1.335 & 1.468 & 1.607 & 1.734 & 1.867 & 2.001 & 2.133 & $ \dots $ \\ \hline
$ l_{opt} $ & 2 & 2 & 2 & 3 & 3 & 3 & 19 & 20 & 20 \\ \hline
\end{tabular}
\end{center}
\label{tb:tab2}
\end{table}

\begin{figure}
\includegraphics{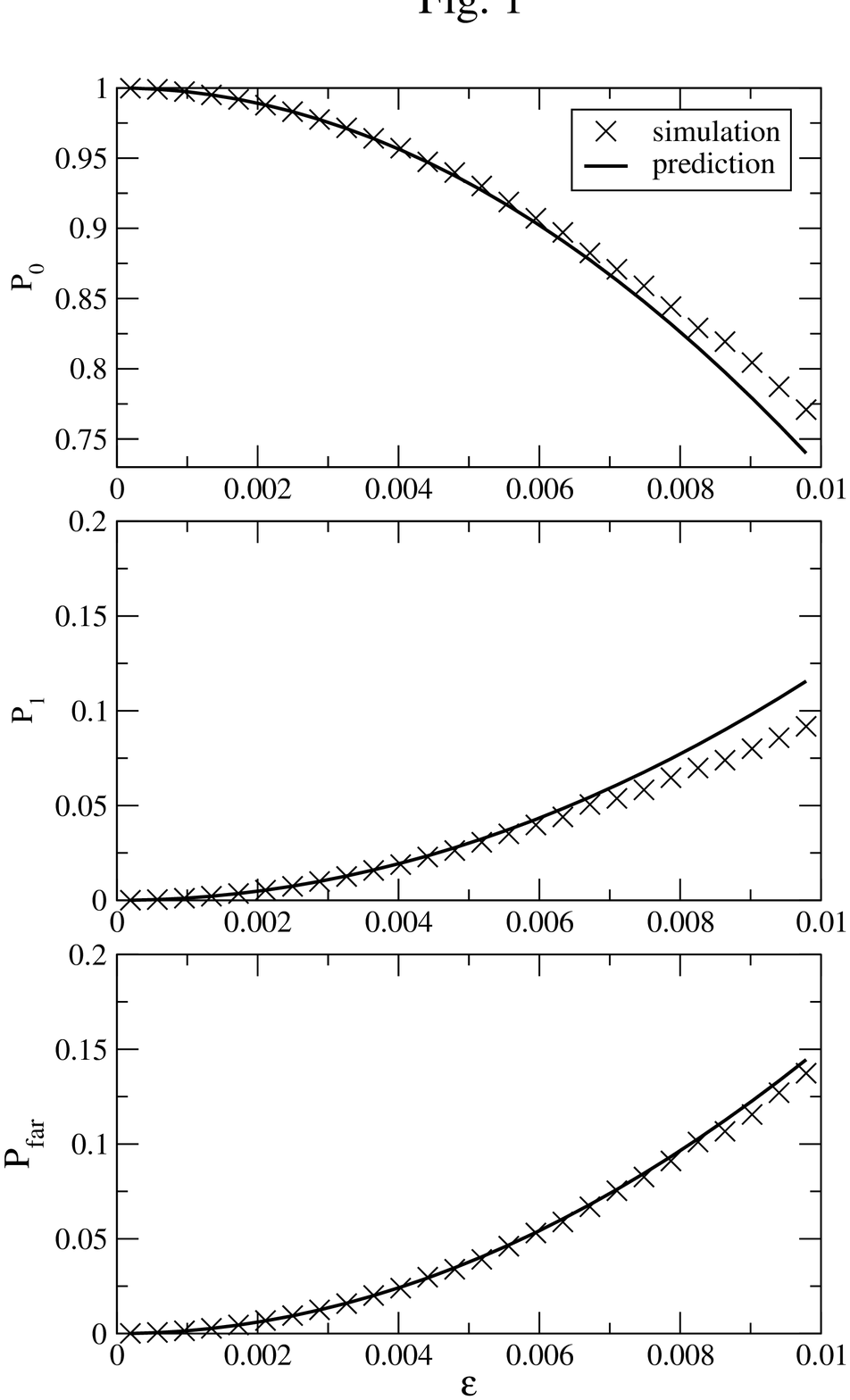}
\caption{
Probability of measurement of the marked state ($P_0$), near states ($P_1$), and far
states ($P_{far}$) as a function of the standard deviation of the noise, $\epsilon$,
given in units of radians. 
Simulation
size is 12 bits, each data point was averaged over 1000 simulation
runs with estimated statistical error of 0.004. X marks show simulated
data, and lines show the predicted values according to Eqs. (\ref{P0}),
(\ref{P1}) and (\ref{Pfar}).
}
\label{fg:fig1}
\end{figure}

\begin{figure}
\includegraphics{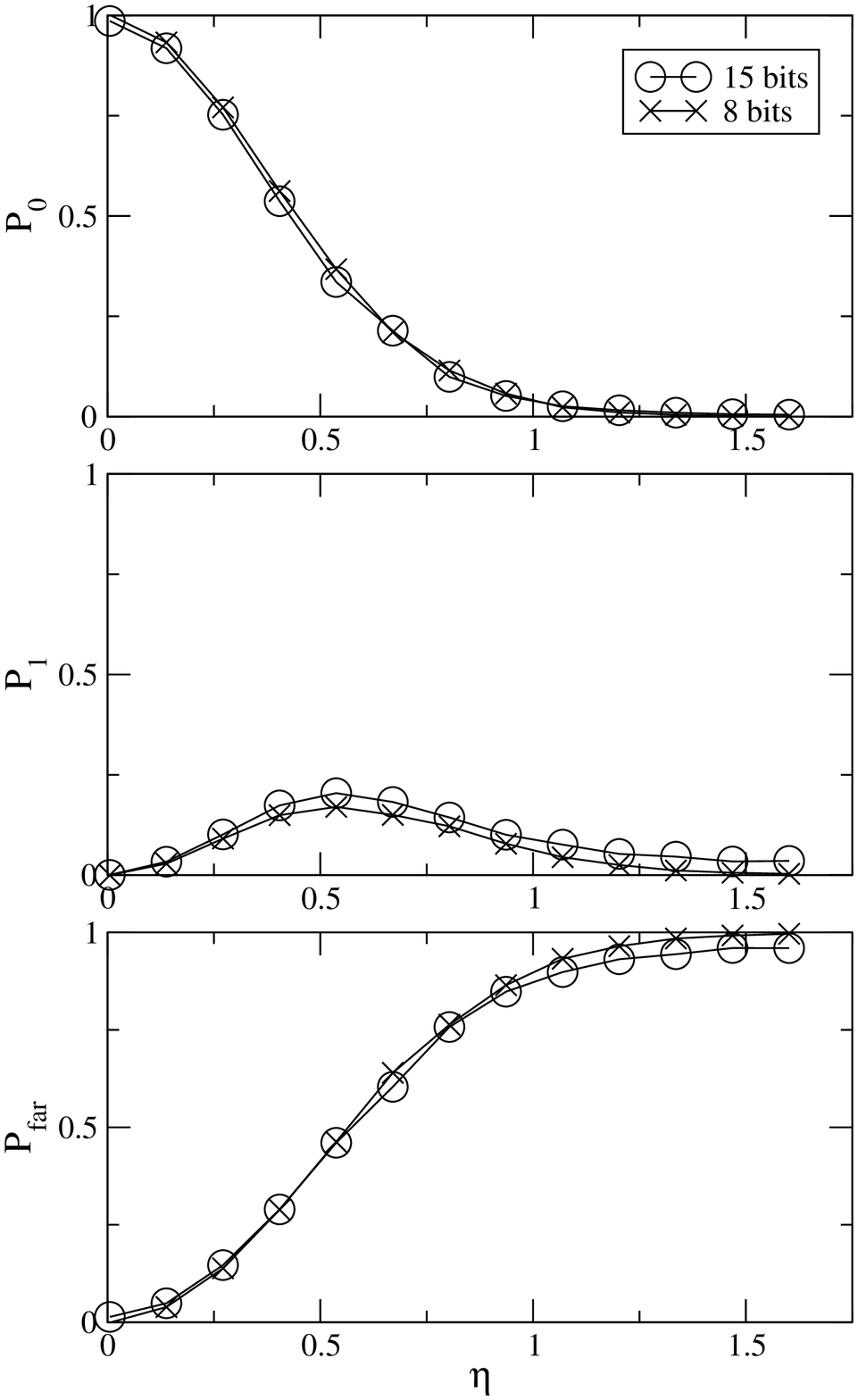}
\caption{
Probability of measurement of the marked state ($P_0$), near states ($P_1$),and far
states ($P_{far}$) as a function of the re-scaled standard deviation of the noise
$\eta = \sqrt{n \sqrt{N}} \epsilon$ (i.e. the standard deviation is given in 
units of $ \frac{1}{\sqrt{n\sqrt{N}}}$). o-marked lines
are results of a 15-bit simulation, and x-marked lines of an 8-bit
simulation. Each data point is averaged over 200 simulation runs.
Estimated statistical error is 0.007.
}
\label{fg:fig2}
\end{figure}

\begin{figure}
\includegraphics{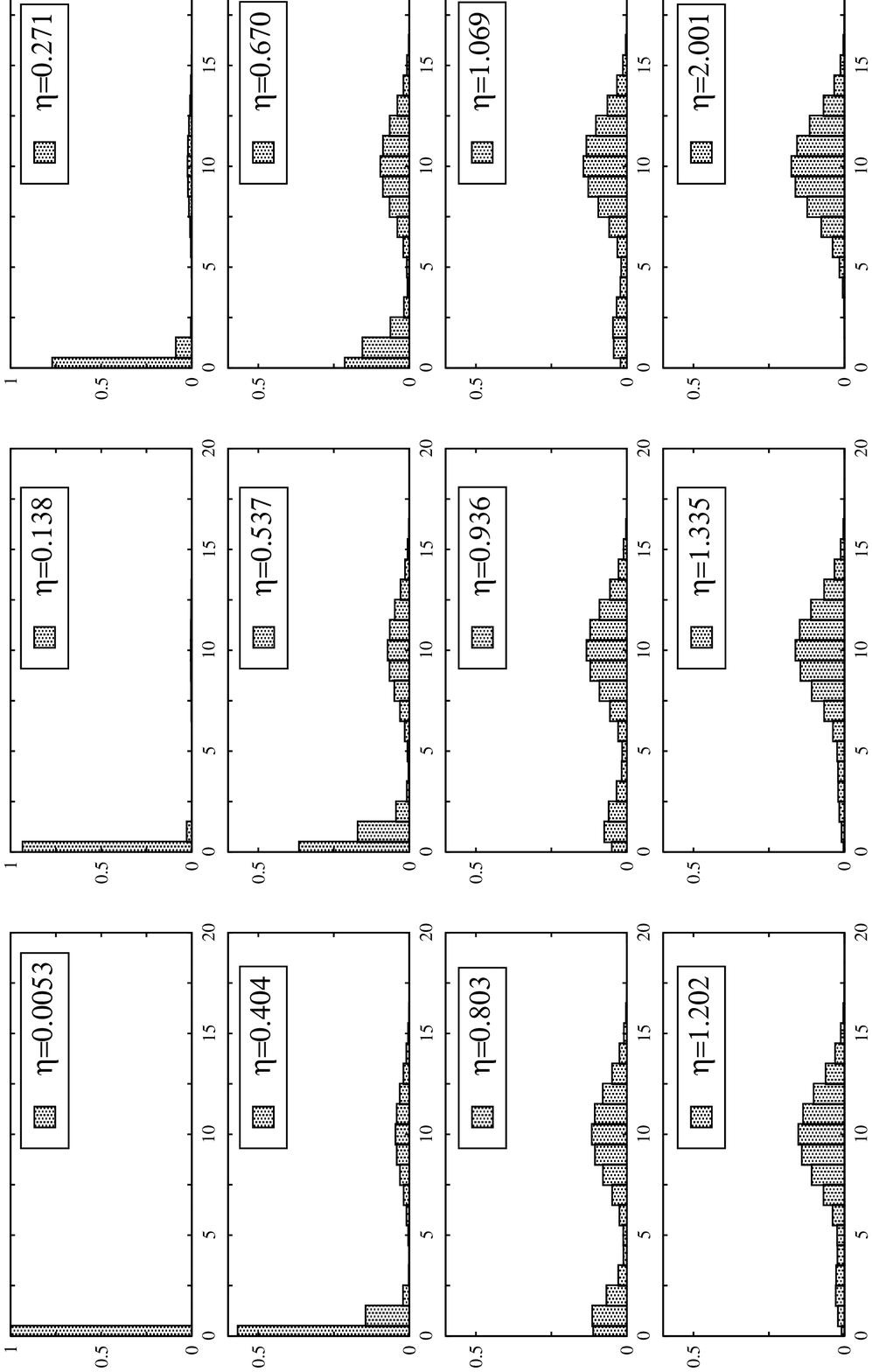}
\caption{
Distribution of probability $P_l$ (y-axis) among the neighborhood classes $l$ (x-axis) for 
($ n=20 $) under different
levels of re-scaled standard deviation of the noise ($\eta = \sqrt{n \sqrt{N}} \epsilon$).
The probabilities in each case are averaged over 300 simulation runs. Estimated statistical error is 
0.005.
}
\label{fg:fig3}
\end{figure}

\begin{figure}
\includegraphics{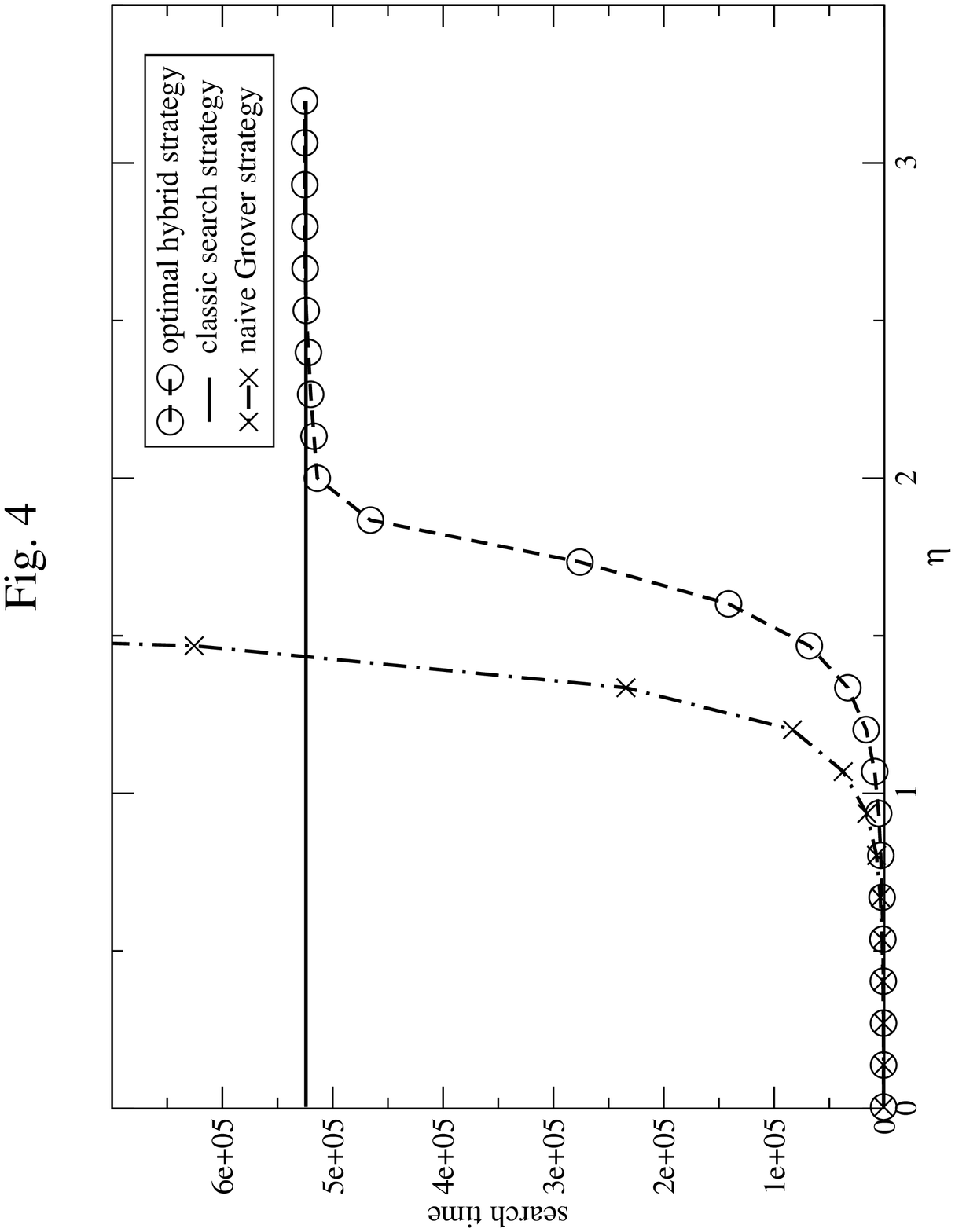}
\caption{
Time of search for different strategies as a function of the re-scaled standard deviation 
of the noise $\eta = \sqrt{n \sqrt{N}} \epsilon$. The data is that of the simulations
used to produce Fig. \ref{fg:fig3} (number of bits is 20). 
Time is measured in units of classical computation time, assuming  $ \tau _{q}=1 $.
The x- marked line denotes the averaged searching time curve of the naive quantum search. 
The o- marked line denotes the averaged searching time curve of the optimal hybrid strategy. 
The solid line, is the averaged searching time boundary determined by the averaged classical 
searching time (with the value of $\frac{1}{2} N$), above which any other search strategy 
is inefficient. 
}
\label{fg:fig4}
\end{figure}

\end{document}